\documentclass[acmsmall,screen]{acmart}

\usepackage{enumitem}
\usepackage{multirow}
\usepackage{algorithm}
\usepackage{algorithmic}
\usepackage{colortbl}
\setlist[enumerate]{align=left}
\AtBeginDocument{%
  }

\setcopyright{none}
\acmJournal{PACMPL}
\acmVolume{11}
\acmNumber{OOPSLA1}
\acmArticle{1}
\acmYear{2027}
\acmMonth{4}
\begin{document}

\title{VeruSteer: Learning What to Repair Next in Verus}

\author{Chao Jiang}
\orcid{0009-0008-3173-5196}
\affiliation{%
  \institution{Shenzhen University}
  \department{College of Computer Science and Software Engineering}
  \city{Shenzhen}
  \country{China}
}
\email{chaoccc22@gmail.com}

\author{Ding Wang}
\orcid{0009-0008-2582-5224}
\affiliation{%
  \institution{Shenzhen University}
  \department{College of Computer Science and Software Engineering}
  \city{Shenzhen}
  \country{China}
}
\email{lifetruthuuu@gmail.com}

\author{Chaohua Yang}
\orcid{0009-0006-8555-2841}
\affiliation{%
  \institution{Shenzhen University}
  \department{College of Computer Science and Software Engineering}
  \city{Shenzhen}
  \country{China}
}
\email{chaohua.ych@gmail.com}

\author{Dugang Liu}
\orcid{0000-0003-3612-709X}
\correspondingauthor
\affiliation{%
  \institution{Shenzhen University}
  \department{College of Computer Science and Software Engineering}
  \city{Shenzhen}
  \country{China}
}
\email{dugang.ldg@gmail.com}

\author{Zhiwu Xu}
\orcid{0000-0001-6727-440X}
\affiliation{%
  \institution{Shenzhen University}
  \department{College of Computer Science and Software Engineering}
  \city{Shenzhen}
  \country{China}
}
\email{xuzhiwu@szu.edu.cn}

\author{Cheng Wen}
\orcid{0000-0003-1826-6213}
\affiliation{%
  \institution{Shenzhen University}
  \department{College of Computer Science and Software Engineering}
  \city{Shenzhen}
  \country{China}
}
\email{wencheng@xidian.edu.cn}

\author{Zhong Ming}
\orcid{0000-0001-9310-3460}
\affiliation{%
  \institution{Shenzhen University}
  \department{College of Computer Science and Software Engineering}
  \city{Shenzhen}
  \country{China}
}
\email{mingz@szu.edu.cn}

\author{Yuwei Liu}
\orcid{0000-0001-5170-3388}
\affiliation{%
  \institution{Ant Group}
  \department{Ant Research}
  \city{Hangzhou}
  \country{China}
}
\email{lyw458372@antgroup.com}

\author{Xinyi Wan}
\orcid{0009-0004-3543-499X}
\affiliation{%
  \institution{Ant Group}
  \department{Ant Research}
  \city{Hangzhou}
  \country{China}
}
\email{wanxinyi.wxy@antgroup.com}

\author{Lin Huang}
\orcid{0009-0002-5659-1471}
\affiliation{%
  \institution{Ant Group}
  \department{Ant Research}
  \city{Hangzhou}
  \country{China}
}
\email{linyu.hl@antgroup.com}

\renewcommand{\shortauthors}{D. Wang, C. Jiang, D. Liu, Z. Xu, C. Wen, Z. Ming, Y. Liu, X. Wan, and L. Huang}

\begin{abstract}\label{sec:abstract}
LLM-based verification systems repair Rust proofs by revising programs in response to Verus diagnostics.
When several error types are reported, pipelines such as AutoVerus and VeruSAGE select the next repair target using a preset priority.
Yet each repair changes the program and the errors that follow, so the same reported types can require different choices at different states.
We present VeruSteer, which learns to select error types from the outcomes of alternative repair trajectories.
It constructs repair trees offline and compares paths from the same initial program by verification success, repair rounds, and final verification progress.
These comparisons supervise a selector that combines step-wise imitation with trajectory preference ranking.
The selector uses self-attention over the program, verifier feedback, and earlier selections to rank the currently reported types.
It replaces the default priority within a single repair trajectory, keeping the existing repair agents and adding no repair or verifier calls for selection.
We evaluate VeruSteer by replaying its choices on stored AutoVerus and VeruSAGE test trees under four generation configurations.
Across 1,270 test trees, VeruSteer increases verified outcomes from 185 to 194 (14.57\% to 15.28\%), gaining 12 successes and losing three for a net improvement of 0.71 percentage points.
Among trees where the default fails but a recorded alternative verifies, VeruSteer succeeds on 3 of 4 AutoVerus trees and 9 of 11 VeruSAGE trees.
Among trees with recorded alternatives where the default verifies, it retains success on 14 of 15 AutoVerus trees and 12 of 14 VeruSAGE trees.
It also finds shorter successful trajectories and improves final progress on some trees where all recorded trajectories fail.
These replay results provide evidence that learning repair order from trajectory outcomes can yield a modest overall verification gain with existing repair agents.
\end{abstract}



\maketitle

\section{Introduction}
\label{sec:introduction}


Verus~\cite{Verus} verifies Rust~\cite{Rust} programs against formal specifications, using proof annotations to help an SMT solver discharge the resulting obligations.
A verification attempt may fail because the program has compilation or type errors, or because Verus cannot prove an obligation such as an assertion, a loop invariant, or an arithmetic-safety condition.
An unproved obligation does not by itself establish a program bug: the program may instead lack the annotations needed to justify it.
LLM-based systems such as AutoVerus~\cite{AutoVerus} and VeruSAGE~\cite{VeruSAGE} address these failures through iterative repair.
Repair can preserve the existing implementation, specifications, and useful proof annotations while using verifier diagnostics to focus revisions on unresolved obligations, rather than regenerating the entire annotated program.


When several diagnostics are reported together, a repair pipeline must decide which one to address next.
AutoVerus and VeruSAGE classify verifier diagnostics into categories, which we call error types, and invoke a specialized repair agent for the selected type.
They make this selection using a preset priority over error types, falling back to diagnostic order when none of the priority-listed types is present.
This rule does not consider the program, diagnostic details, or earlier repair attempts.
These signals can matter: the program determines how proof obligations depend on one another, diagnostics identify the conditions that remain unproved, and the repair history records which changes have already been tried and what they achieved.

The consequences of target selection extend beyond the current round.
In our motivating example, the loop is annotated with the invariant
\texttt{sum == arithmetic\_sum(i as nat)}; its body increments
\texttt{i} and adds it to \texttt{sum}.
Preserving this invariant requires establishing that the old sum plus
the incremented index equals \texttt{arithmetic\_sum} at the new index.
Together with a bound on this mathematical sum, the same relation
can justify that the machine-integer addition does not overflow.
The invariant failure therefore identifies a proof relation useful
to both obligations, making it a promising first repair target
in this example.
In the recorded tree, the invariant-first branch reaches verification
after subsequent assertion and overflow repairs, whereas the
overflow-first branch leaves an assertion unproved through the
remaining budget (Section~\ref{sec:motivating-example}).
Moreover, later states on the successful trajectory report the same
error types but require different selections, so no fixed ordering
reproduces that trajectory.

We study how often such differences arise in 3,224 repair trees over 684 tasks, collected with AutoVerus and VeruSAGE (Section~\ref{sec:motivation-study}).
A repair tree records alternative repair sequences from one initial program: nodes are program states with verifier feedback, branches record repairs targeting different reported error types, and each root-to-leaf path is a repair trajectory.
Branches within a tree use the same repair configuration and budget.
Of the 718 trees with alternative trajectories, 32.0\% contain an alternative that verifies where the default fails, verifies in fewer rounds, or ends with greater progress when all trajectories fail.
Among states with a uniquely best type under the recorded outcomes, choosing the most frequently best type for each reported set recovers that choice at only 62.0\% of states.
These observations motivate selecting repair targets from their context and judging choices by the trajectories they produce.

Adaptive target selection already appears in StarVerus~\cite{jiang2026starverus}, whose LLM planner analyzes the program and diagnostics to choose a repairer.
This establishes a precedent for contextual selection, but leaves open how to learn such a policy explicitly from observed repair outcomes.
We investigate supervision derived from comparisons of alternative repair trajectories, rather than relying on a prompted planner's analysis alone.
An ordinary run observes the result of its selected target but not where other targets would have led, and the value of a selection may emerge only after several further repairs.
Comparing complete trajectories from the same starting program provides evidence about this delayed effect.

We present VeruSteer, a framework for learning error-type selection from repair trajectories.
During offline data construction, VeruSteer branches on the reported error types, using the pipeline's existing agents and acceptance checks to record one repair outcome for each type.
The resulting paths expose different repair orders under the same budget.
We derive preferences from whether a path verifies, how many rounds a successful path needs, and how much verification progress an unsuccessful path makes.
Comparisons within each tree let both successful and failed paths contribute supervision.

VeruSteer formulates this selection as next-item prediction in sequential recommendation: a trajectory forms the interaction sequence, and error types are the candidate items.
Its error type repair selector uses a self-attention encoder over the current program, verifier feedback, and repair history to score the types available at each state.
It learns through two objectives: step-wise imitation of preferred trajectories and pairwise ranking based on their outcomes.
At deployment, one ranking chooses the next target and the underlying pipeline performs the repair.
This follows a single trajectory under the original budget and adds no repair or verifier call for selection.

We evaluate VeruSteer by replaying its choices on stored test trees from both pipelines, with two generation LLMs and two prompting settings.
We measure whether it retains outcomes already achieved by the default rule and finds better alternatives when the trees contain them.
Where the default fails but an alternative verifies, VeruSteer recovers 75.0\% of the available verification improvements on AutoVerus and 81.8\% on VeruSAGE.
On trees with alternative trajectories where the default verifies, it preserves 93.3\% and 85.7\% of those successes, respectively.
It also finds shorter verified trajectories and improves final progress on trees where no trajectory verifies.

On VeruSAGE, replacing the selector with Random or rules learned from error types alone reduces recovery of verification opportunities to 43.6--54.5\%.
Ablations further show that the program and diagnostics, both training losses, and the representations of earlier selections and state order contribute to this recovery.

This paper makes the following contributions:
\begin{itemize}[leftmargin=1.6em, labelsep=0.5em, itemsep=1pt, topsep=2pt, beginpenalty=10000]
    \item \textbf{Characterizing error selection.} We formulate the choice of repair target in terms of trajectory outcomes, and construct a corpus of repair trees that shows both the opportunities missed by preset priorities and how the better choice depends on context.
    \item \textbf{Learning from alternative trajectories.} We develop VeruSteer, which combines preferences from successful and failed repair paths with a sequence model that selects among the currently reported error types of an existing repair pipeline.
    \item \textbf{Controlled evaluation of repair selection.} We compare selectors over the same recorded transitions in AutoVerus and VeruSAGE, measuring default-outcome preservation and alternative recovery, and examine selection strategies, model components, and contextual inputs.
\end{itemize}

\section{Background}
\label{sec:background}

\subsection{Verus}
\label{sec:background-verus}

Verus~\cite{Verus} is a deductive verifier for Rust~\cite{Rust}: developers
annotate functions with specifications and proof code, and Verus discharges
the resulting verification conditions with the SMT solver Z3~\cite{de2008z3}.
A function's contract consists of \texttt{requires} and \texttt{ensures}
clauses, which may call ghost \texttt{spec} functions.
Proof code, such as loop
\texttt{invariant} and \texttt{decreases} clauses, assertions, and
\texttt{proof} lemmas, helps Verus discharge the obligations that the contract
and the implementation induce, and is erased before compilation.
Before attempting any proof, Verus checks that the program is well-typed and
follows Verus's language rules, such as the separation of ghost and executable
code.
Each failed check yields a diagnostic, which consists of a message naming the
kind of failure (e.g., \texttt{invariant not satisfied at end of loop body}), a
source location, and the offending expression.
A single run can report several diagnostics, followed by a summary line such
as \texttt{verification results:: 5 verified, 2 errors}.
The line counts the verified proof obligations and the reported errors, and a
program verifies when it reports no errors.

\subsection{Error Selection in LLM-Based Repair}
\label{sec:background-pipelines}

Writing specifications and proof code by hand takes considerable expertise.
LLM-assisted verification systems for Verus automate this work and refine
their output in a loop with the verifier.
Each round runs Verus on the current program, asks an LLM to edit the program
based on the diagnostics, and checks the resulting candidates, until the
program verifies, or a round budget runs out.
A round therefore involves three decisions: which failure to target, how to
repair it, and which candidate to carry forward.
Retrieved repository context~\cite{RagVerus} and
counterexamples~\cite{ExVerus} help the LLM understand a failure and
construct a repair, while tree search~\cite{AlphaVerus} and candidate
ranking~\cite{ExVerus} decide which candidates to carry forward.
The first decision determines which failure the repair round addresses.

Existing systems make this decision in different ways.
SAFE's self-debugging~\cite{SAFE} and AlphaVerus's
Treefinement~\cite{AlphaVerus} pass the program and all of its diagnostics to
the model and leave the target implicit.
AutoVerus~\cite{AutoVerus}, VeruSAGE~\cite{VeruSAGE}, and
VeriStruct~\cite{VeriStruct} make the choice explicit: they select one
diagnostic and dispatch a repair agent or module specialized for its class.
Specialization lets each prompt carry class-specific repair guidance.
AutoVerus's overflow agent, for instance, asks for loop invariants that bound
every variable in the overflowing expression.
All three systems make the choice with a preset priority over error types.
StarVerus~\cite{jiang2026starverus} instead lets an LLM planner make it.
Given the current implementation, specifications, diagnostics, and project
context, the planner diagnoses a root cause and proposes a repair order over
error types, and the first type in that order selects a specialized repairer.
The planner's choice thus adapts to the program and its diagnostics.
These approaches do not explicitly learn preferences among error types from
comparisons of the downstream outcomes of alternative repair orders.

\begin{figure}[t]
    \centering
    \includegraphics[width=0.8\linewidth]{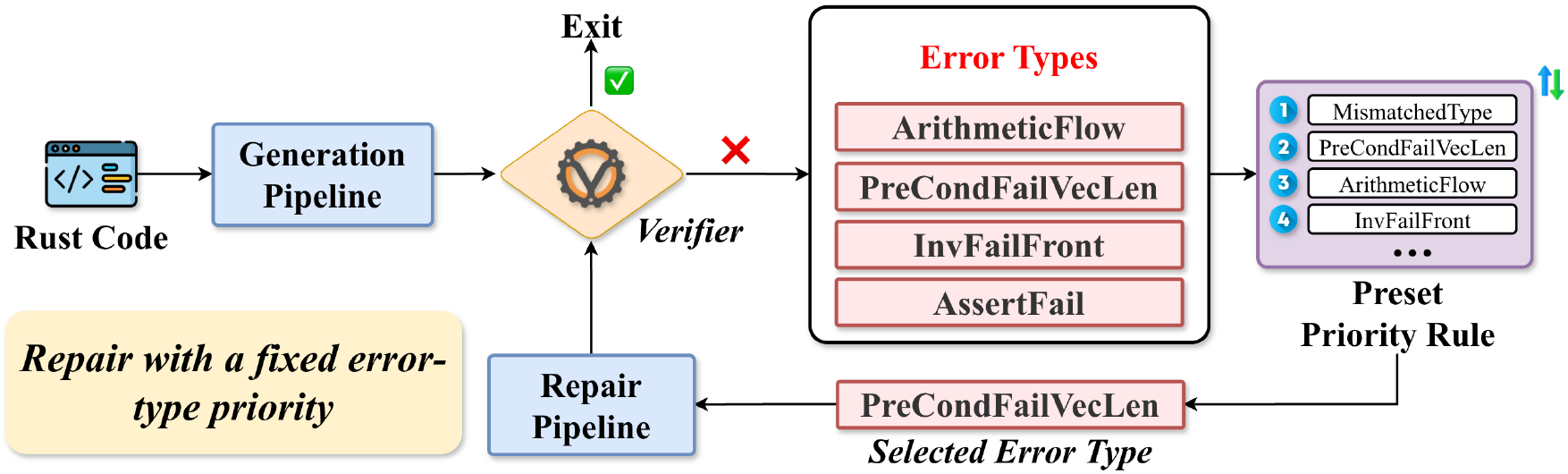}
    \caption{A repair loop with fixed error-type priorities.}
    \Description{Rust code passes through a generation pipeline and a verifier.
    Successful verification ends the loop. On failure, a fixed priority rule
    selects an error type from the reported diagnostics. The example selects
    PreCondFailVecLen, which is passed to the repair pipeline before the program
    is checked again.}
    \label{fig:background}
    \vspace{-10pt}
\end{figure}

In AutoVerus and VeruSAGE, each round classifies diagnostics by error type,
selects one type, and invokes the corresponding repair agent
(Figure~\ref{fig:background}).
The agent uses the selected diagnostic to decide how to edit the program.
If the patch passes the pipeline's acceptance checks, the next round starts
from the patched program.
Otherwise the round is spent, and the next one starts from the same program.

Both pipelines implement their preset priority as a short hard-coded list,
which applies whenever Verus reports several types at once.
The list ranks \texttt{MismatchedType} (type mismatches) first, then
\texttt{PreCondFailVecLen} (vector-length preconditions),
\texttt{ArithmeticFlow} (possible arithmetic overflow), and
\texttt{InvFailFront} (invariants failing before the loop).
VeruSAGE appends
\texttt{InvFailEnd} (invariants failing at the end of the loop body).
If none of these types is present, the first diagnostic in Verus's output is
repaired.
Type mismatches come first because they keep Verus from attempting any proof.
Verus imposes no such order on proof failures, so the list fixes one in
advance, the same for every program and every round.
The rule thus depends only on which types are present and on the order in
which Verus prints its diagnostics.

\section{Motivation}
\label{sec:motivation}

\subsection{A Motivating Example}
\label{sec:motivating-example}

\begin{figure}[t]
    \centering
    \includegraphics[width=1.0\linewidth]{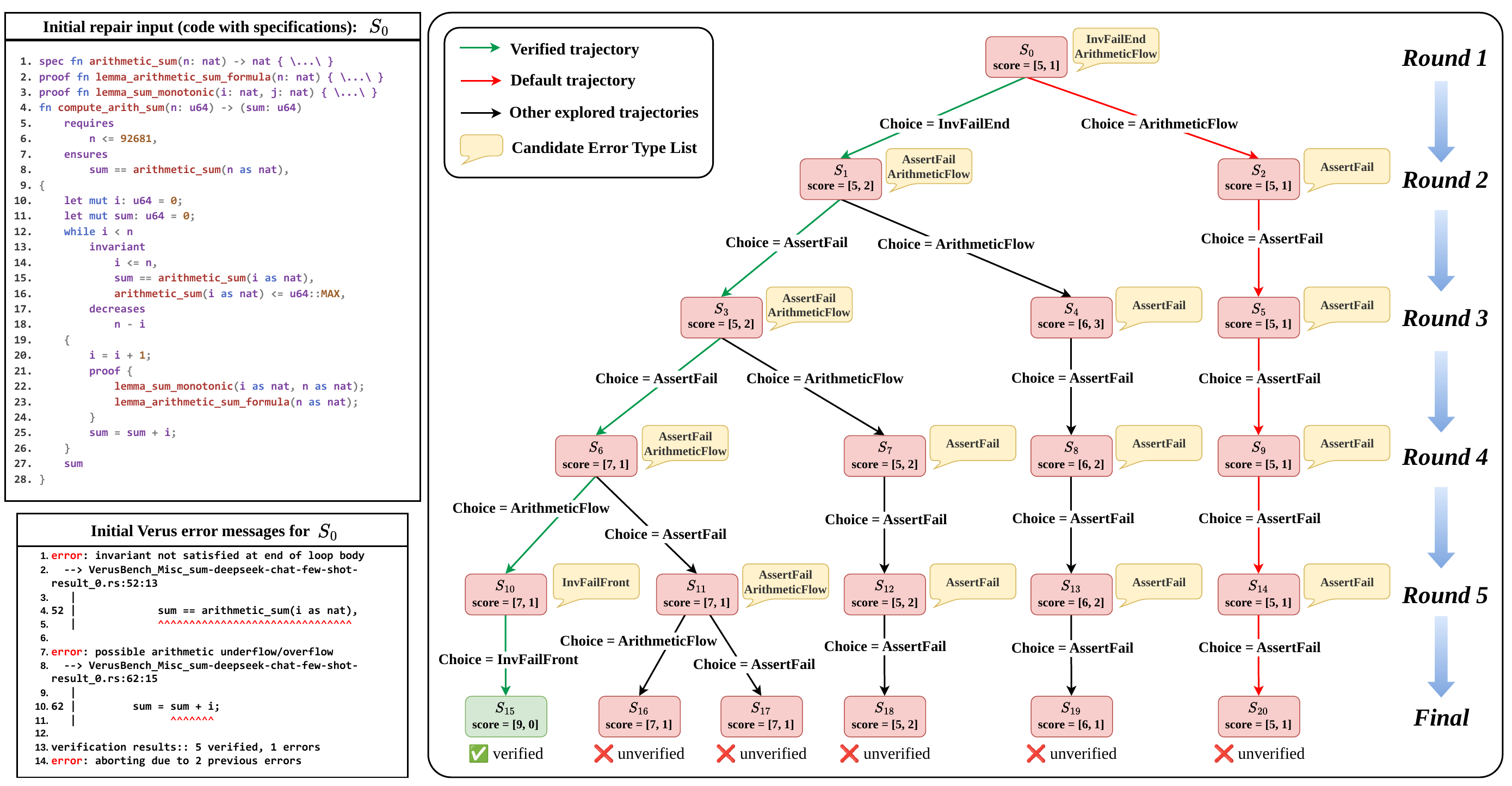}
    \caption{Repair tree of the motivating example.}
    \Description{A repair tree in which two repair orders from the same initial failing state lead to a failed state and a verified state, respectively.}
    \label{fig:insight_case}
    \vspace{-10pt}
\end{figure}

Consider \texttt{compute\_arith\_sum}, which sums $1+2+\cdots+n$ in a
 loop using \texttt{u64} arithmetic.
Its loop invariants tie the running sum to the specification function
\texttt{arithmetic\_sum} and bound it by \texttt{u64::MAX}.
In the initial state $S_0$, Verus reports (see the error messages at the lower
left of Figure~\ref{fig:insight_case}) that a loop invariant may fail at the
end of the loop body (\texttt{InvFailEnd}) and that \texttt{sum = sum + i} may
overflow (\texttt{ArithmeticFlow}).
The default rule (Section~\ref{sec:background-pipelines}) ranks
\texttt{ArithmeticFlow} above \texttt{InvFailEnd}, so it repairs the
overflow first.
Figure~\ref{fig:insight_case} shows the repair tree that VeruSAGE with
DeepSeek-Chat (five-shot) produces for this program within a budget of five
rounds.
Besides the default trajectory, the tree contains the trajectories that
select a different reported type at some state.
Each node's score $[x,y]$ comes from the summary line of Verus's output,
where $x$ is the number of verified proof obligations and $y$ the number of
reported errors.
For $S_0$, the summary line \texttt{5 verified, 1 errors} gives $[5,1]$.
The error count $y$ is not the number of error types: $S_0$ counts one error,
yet its diagnostics fall into two types, \texttt{InvFailEnd} and
\texttt{ArithmeticFlow}, which are listed beside the node and on which the
tree branches.

The default (red) trajectory selects \texttt{ArithmeticFlow} at
$S_0$ and proceeds to $S_2$, where \texttt{AssertFail} is the
only reported type.
It repeatedly targets \texttt{AssertFail} in each of the four remaining rounds and ends
unverified at $S_{20}$.
Its score stays at $[5, 1]$ from $S_0$ to $S_{20}$, indicating that none of its five
rounds makes measurable progress.
The verified (green) trajectory instead selects \texttt{InvFailEnd} at
$S_0$ and verifies the program at $S_{15}$, with score $[9,0]$, after five
rounds.
The default rule thus fails to identify a successful repair order that fits within the
same budget.

Getting the first choice right is not enough.
Along the verified trajectory, $S_1$, $S_3$, and $S_6$ all report
\{\texttt{AssertFail}, \texttt{ArithmeticFlow}\}.
At $S_1$ and $S_3$, the verified trajectory selects \texttt{AssertFail},
whereas \texttt{ArithmeticFlow} leads only to the unverified leaves
$S_{19}$ and $S_{18}$.
At $S_6$, it selects \texttt{ArithmeticFlow}, whereas \texttt{AssertFail}
leads only to the unverified leaves $S_{16}$ and $S_{17}$.
The default rule, which ranks \texttt{ArithmeticFlow} above
\texttt{AssertFail}, goes wrong at $S_1$ and $S_3$, and the reverse ranking
goes wrong at $S_6$.
No fixed ranking of the two types reproduces the verified trajectory.

We compare scores by preferring larger $x$, breaking ties by smaller $y$
(Section~\ref{sec:trajectory-quality}).
A higher score in the next state does not necessarily lead to a better final
outcome.
At $S_1$, \texttt{ArithmeticFlow} yields $S_4$ with score $[6, 3]$, one more
verified obligation than $S_3$ with $[5, 2]$, yet only \texttt{AssertFail}
leads to verification.

In this tree, the better choice thus varies from state to state, and the
outcomes of the subsequent trajectories reveal which choice it is.

\subsection{Do the Observations Generalize?}
\label{sec:motivation-study}

The example makes two observations on a single task: the default order can
miss a better trajectory, in this case a verified one, and the same set of
reported error types can favor different choices at different states.
We now examine how widely both observations hold in a corpus of repair trees
like the one in Figure~\ref{fig:insight_case}.
Each tree starts from one initial program repaired by AutoVerus or
VeruSAGE.
Whenever a state reports several error types, the tree branches once for
each type (Section~\ref{sec:tree-exploration}).
Its trajectories therefore share the initial program, pipeline, LLM,
prompting setting, and repair budget, and differ in their error-type
selections, so their outcomes can be compared directly.
The corpus contains 3,224 such trees over 684 tasks, collected with two LLMs
and two prompting settings (Section~\ref{sec:setup}).

We first identify trees in which the default trajectory reaches a state with
several reported error types, offering a choice of repair target.
Of the 3,224 trees, 718 (22.3\%) contain such a state, and all of them
record alternative trajectories.
In the other 2,506 trees, every selection state on the recorded default
trajectory offers only one error type.

We compare trajectories by verification first, then by the number of rounds
if both verify, and by the progress of the final state if neither does (more
verified proof obligations, then fewer reported errors).
Section~\ref{sec:trajectory-quality} defines this order formally.
In 230 of the 718 trees (32.0\%), an alternative trajectory is strictly
better than the default.
This share is similar for AutoVerus (32.7\%) and VeruSAGE (31.2\%).
Of these 230 trees, 67 (9.3\% of the 718) show gains in verification or
repair efficiency.
In 39 of them, as in the example, an alternative verifies a program that the
default leaves unverified.
In the other 28, both verify, but an alternative needs fewer rounds.
In the remaining 163, no trajectory verifies, and an alternative ends with
more progress.

We next test the second observation, that the same reported types can favor
different choices at different states.
A fixed ranking always picks the same type for the same reported types, so it
cannot follow such variation.
In the example, the better choice for
\{\texttt{AssertFail}, \texttt{ArithmeticFlow}\} was \texttt{AssertFail} at
$S_1$ and $S_3$ but \texttt{ArithmeticFlow} at $S_6$.
To measure this across the corpus, we take every state at which a tree tries
several types and follow each type to the best trajectory below it.
At 4,943 such states, one type leads to a strictly better trajectory than all
others, and we call it the best choice.
Among them, the states that report the same pair as the example favor
\texttt{ArithmeticFlow} at 361 and \texttt{AssertFail} at 276.
At 97.0\% of the 4,943 states, another of them reports the same types but has
a different best choice.
Even mapping each set of reported types to its most frequent best choice
recovers the best choice at only 62.0\% of them.
Any fixed ranking maps each set to one type, so it misses the best choice at
38.0\% or more of them.

Both observations of the example thus recur across the corpus.
These results motivate a selection policy that uses the repair context to
choose among reported error types, guided by the outcomes of alternative
repair trajectories.

\section{Problem Formulation}\label{sec:problem-formulation}

This section makes the selection studied in Section~\ref{sec:motivation}
precise: the repair interaction in which it occurs, the order by which its
trajectories are compared, and the information on which it may depend.

\subsection{Repair as Verifier-in-the-Loop Interaction}
\label{sec:repair-interaction}

A repair run starts from an initial program $c_1$ that Verus does not verify,
with a fixed repair pipeline, LLM, prompting setting, and budget of $B$
rounds.
At round $i$, the verifier maps the current program $c_i$ to
\begin{equation}
    (x_i,y_i,e_i)=\mathcal{V}(c_i),
    \label{eq:verifier-step}
\end{equation}
where $x_i$ and $y_i$ are the numbers of verified proof obligations and
reported errors in Verus's summary line
(Section~\ref{sec:motivating-example}), and $e_i$ is the full verifier feedback,
which contains the diagnostics and the summary line.
We call the pair $(c_i,e_i)$ the $i$-th state of the run.
If $c_i$ fails to compile, Verus attempts no proof, and we set
$(x_i,y_i)=(-1,-1)$.
The program verifies when Verus completes successfully and reports $y_i=0$.
Otherwise, the pipeline classifies the diagnostics in $e_i$ using a fixed
vocabulary $\mathcal{T}$ of error types and collects all identified types
into the candidate set:
\begin{equation}
    \mathcal{T}^{(i)}_{\mathrm{cand}}
    =\mathrm{Err}(e_i)\subseteq\mathcal{T}.
    \label{eq:candidate-set}
\end{equation}
If this set is nonempty and budget remains, the pipeline selects a target
type $t_i\in\mathcal{T}^{(i)}_{\mathrm{cand}}$.
It passes $c_i$ and the first diagnostic of type $t_i$ in $e_i$ to the repair
agent for $t_i$, which generates candidate revisions of
$c_i$~\cite{AutoVerus,VeruSAGE}.
We write the resulting repair step as:
\begin{equation}
    c_{i+1}=\operatorname{Repair}(c_i,e_i,t_i),
    \label{eq:repair-transition}
\end{equation}
where $\operatorname{Repair}$ generates and checks the candidates under the
pipeline's own rules, which may also depend on earlier rounds, and returns
the revision it accepts, or $c_i$ if it accepts none.
Each step consumes one round of the budget, including a step that accepts no
candidate and leaves $c_{i+1}=c_i$.
The run stops when the program verifies, when no error type can be extracted
from the verifier feedback, or after $B$ rounds.
A run with $n\leq B$ rounds produces a trajectory denoted by:
\begin{equation}
    \tau=(c_1,e_1,t_1,\ldots,c_n,e_n,t_n,c_{n+1},e_{n+1}),
    \label{eq:trajectory}
\end{equation}
which reduces to $(c_1,e_1)$ when $n=0$.
Its final state $(c_{n+1},e_{n+1})$ comes from checking $c_{n+1}$ with
Equation~(\ref{eq:verifier-step}), and this check does not count as a round.

\subsection{Trajectory Quality}
\label{sec:trajectory-quality}

\begin{figure}[t]
    \centering
    \includegraphics[width=0.95\linewidth]{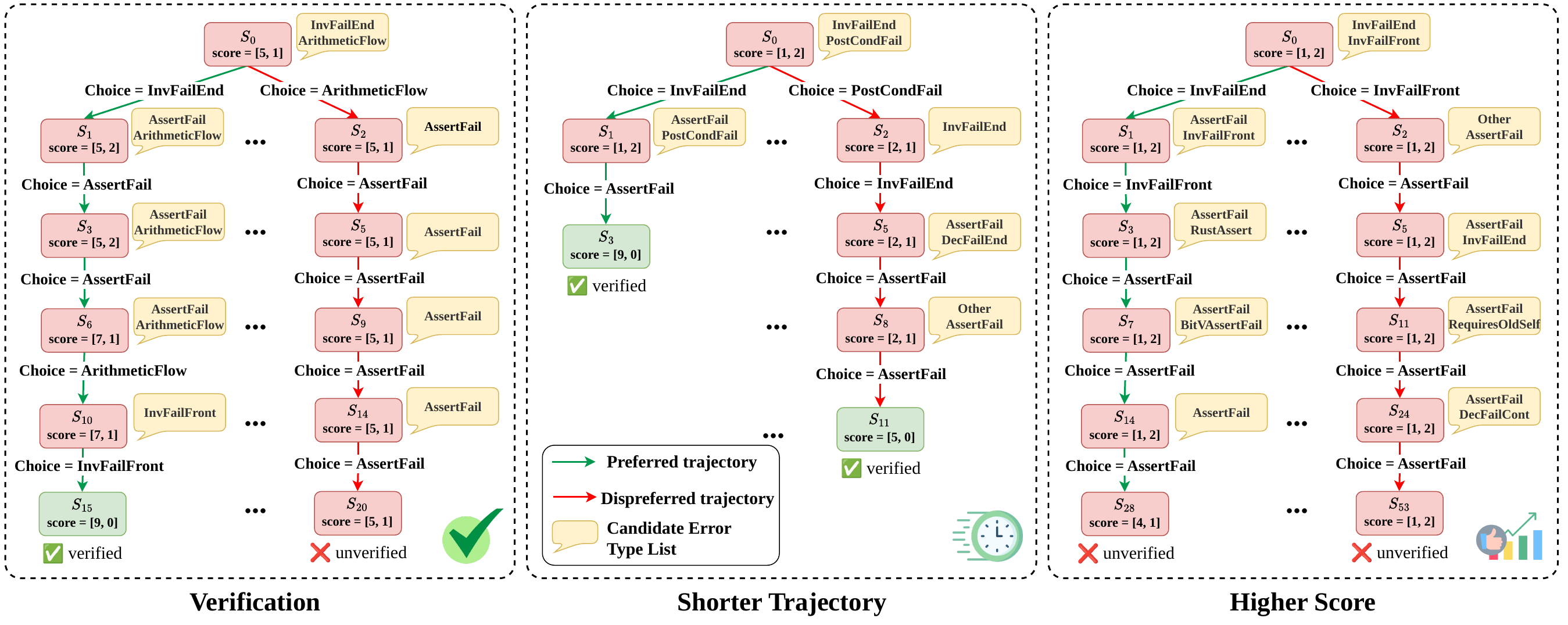}
    \caption{Hierarchical trajectory preference.}
    \Description{Three repair-tree examples illustrate preferences based on
    verification success, successful-trajectory length, and terminal progress.}
    \label{fig:three_better_defin}
    \vspace{-10pt}
\end{figure}

We compare the trajectories of runs with the same setup, that is, the same
initial program, pipeline, LLM, prompting setting, and budget.
Their outcomes differ in three respects: whether the final program verifies,
how many rounds the run takes, and how much progress the final state makes.
For a trajectory $\tau$ with $n$ rounds, let $V(\tau)\in\{0,1\}$ indicate
whether its final program verifies.
The progress score of a state is the pair $Q(c_i,e_i):=(x_i,y_i)$, the score
$[x,y]$ of Section~\ref{sec:motivating-example}.
We compare progress scores by prioritizing more verified proof obligations,
breaking ties by fewer reported errors:
$(x,y)>(x',y')$ if and only if
$x>x'$ or $(x=x' \land y<y')$.
We write $Q(\tau):=Q(c_{n+1},e_{n+1})$ for the progress of the final state of
$\tau$.
For trajectories $\tau$ and $\tau'$ with $n$ and $n'$ rounds, the trajectory preference relation $\succ$, where $\tau\succ\tau'$ reads ``$\tau$ is better than $\tau'$'', is expressed as:
\begin{equation}
    \tau\succ\tau'
    \iff
    \begin{cases}
        V(\tau)>V(\tau'), & \text{if }V(\tau)\neq V(\tau'),\\
        n<n', & \text{if }V(\tau)=V(\tau')=1,\\
        Q(\tau)>Q(\tau'), & \text{if }V(\tau)=V(\tau')=0.
    \end{cases}
    \label{eq:trajectory-preference}
\end{equation}
Verification comes first, since it is the goal of repair.
Between two verified trajectories, the one with fewer repair rounds is
preferred.
Between two unverified ones, the one whose final state makes more progress is
better.
Figure~\ref{fig:three_better_defin} illustrates the three cases.

Two trajectories are tied when neither is better, for example, when both
verify in the same number of rounds.
A trajectory is best if no trajectory of a run with the same setup is better.
Because the LLM samples the candidates, the same state and target type can
lead to different successors, so exhaustively exploring these trajectories
is impractical.
We therefore compare trajectories recorded in repair trees such as the one
in Figure~\ref{fig:insight_case}.
Each expanded node samples one successor per candidate type
(Section~\ref{sec:tree-exploration}), so the recorded root-to-leaf paths form
a finite sample of the possible trajectories within budget $B$.

\subsection{The Error-Selection Problem}
\label{sec:selection-problem}

The pipeline selects $t_i$ after Verus checks $c_i$ and before the repair
agent runs.
At that point it knows the $i$-th state $(c_i,e_i)$ and the repair history
$\mathcal{H}_i:=(c_1,e_1,t_1,\ldots,c_{i-1},e_{i-1},t_{i-1})$ of earlier
states and the types selected in them.
We call $\mathcal{C}_i:=(c_i,e_i,\mathcal{H}_i)$ the decision context.
It covers the trajectory up to the $i$-th state, while the consequences of
the choice appear only in later states.
A selection policy $\pi$ maps each decision context to one of its candidate
types, $t_i=\pi(\mathcal{C}_i)\in\mathcal{T}^{(i)}_{\mathrm{cand}}$.
The default rule of Section~\ref{sec:background-pipelines} is one such
policy.
It decides from the candidate set and, when no priority-listed type is
reported, from the order of the diagnostics.
The trajectory it induces is the default trajectory $\tau^0$.

\noindent\textbf{Problem 1 (Trajectory-Aware Error Selection).}
Given a repair pipeline and repair trees collected from the training instances, learn a selection policy \(\pi\) that generalizes to held-out instances by selecting the most appropriate error type for each repair state, thereby producing trajectories with outcomes that are as favorable as possible under \(\succ\).

\section{The VeruSteer Framework}
\label{sec:framework}

\begin{figure}[t]
    \centering
    \includegraphics[width=0.95\linewidth]{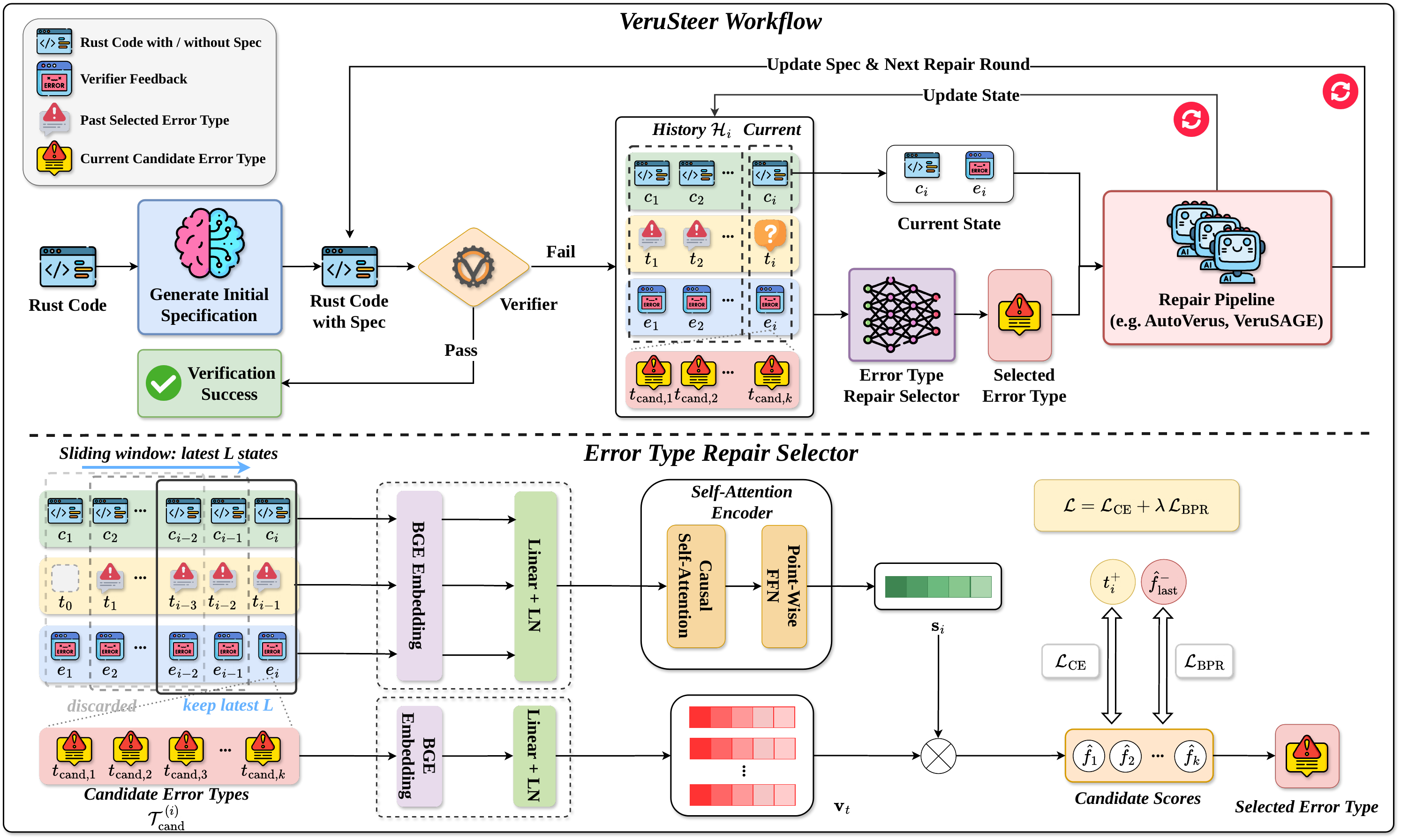}
    \caption{Overview of VeruSteer.}
    \Description{VeruSteer uses offline trajectory supervision to train an
    error type repair selector that controls an otherwise unchanged repair pipeline.}
    \label{fig:workflow}
    \vspace{-10pt}
\end{figure}

Section~\ref{sec:problem-formulation} defined the decision context $\mathcal{C}_i$, the candidate set $\mathcal{T}^{(i)}_{\mathrm{cand}}$, and the trajectory preference $\succ$, and Problem~1 asks for a selection policy $\pi$ learned from recorded repair trees.
VeruSteer expands repair trees to obtain alternative trajectories from the same initial program, turns their outcomes under $\succ$ into supervision, and trains an error type repair selector to determine which error type should be prioritized for repair in the current decision context (Figure~\ref{fig:workflow}).
At deployment, this model replaces the default rule and leaves the rest of the pipeline unchanged.

\subsection{Repair-Tree Construction}
\label{sec:tree-exploration}

A normal run selects one type at each state, so it never observes where the other candidate types would have led.
To observe this, we expand a repair tree $\mathcal{G}$ from each initial program.
At each state $(c_i,e_i)$, instead of selecting one type, we run the repair step $\operatorname{Repair}(c_i,e_i,t)$ once for every candidate type $t\in\mathcal{T}^{(i)}_{\mathrm{cand}}$, with the pipeline's own agents and acceptance checks.
Each step yields a child state, which we expand in the same way.
As in a normal run, a step that accepts no candidate yields a child with the parent's program.
A node becomes a leaf when its program verifies, when its feedback yields no candidate type, or when $B$ rounds have been spent.
We set $B=5$.
The underlying pipelines realize most of their repair capability within five rounds, and the budget keeps expansion affordable.
Figure~\ref{fig:insight_case} shows such a tree: at $S_0$, it branches once for \texttt{InvFailEnd} and once for \texttt{ArithmeticFlow}.

Each node holds a state with its round, candidate set, and progress score, and each edge holds the type selected at its parent.
A root-to-leaf path is therefore a trajectory, and its part up to a node is the decision context of the selection made at that node.
Each path provides the decision contexts and selected types along it, together with the outcome of the resulting trajectory.

The path that follows the default rule at every node is the default trajectory $\tau^0$, and the other paths are alternative trajectories.
All of them start from the same initial program under the same setup, so $\succ$ compares them directly.

\subsection{Preference Supervision}
\label{sec:preference-supervision}

The trees supervise the selector through comparisons between trajectories of the same tree.
Each comparison is an ordered pair $(\tau^+,\tau^-)$ with $\tau^+\succ\tau^-$.
The preferred trajectory $\tau^+$ verifies where the dispreferred trajectory $\tau^-$ does not, verifies in fewer rounds, or, when neither verifies, ends with more progress.

We form the pairs of every tree by the same rule, which follows the three cases of $\succ$.
A verified trajectory is paired with the unverified trajectories and with the slower verified ones.
When no trajectory of the tree verifies, a trajectory with the most progress is paired with those that end with less.
A tree in which no trajectory beats another, such as a tree with a single path, yields no pair.
Whether a trajectory follows the default rule plays no part: $\tau^0$ is paired like any other trajectory.
For example, the tree in Figure~\ref{fig:insight_case} has one verified trajectory, which ends at $S_{15}$, and five unverified ones, one of which is the default trajectory.
The verified trajectory is $\tau^+$, and each unverified one is a $\tau^-$.
Each tree contributes at most 1,000 pairs, and we denote the resulting set by $\mathcal{D}_{\mathrm{pair}}$.

Each pair carries two kinds of information.
The preferred trajectory $\tau^+$ shows which type was selected in each of its decision contexts.
The pair as a whole shows which of two trajectories from the same initial program ends better.
Section~\ref{sec:training-deployment} trains the selector on both.

\subsection{Error Type Repair Selector}
\label{sec:selector}

The selector implements the policy $\pi$ of Problem~1.
Given a decision context $\mathcal{C}_i$, it scores the error types and selects the highest-scoring type in $\mathcal{T}^{(i)}_{\mathrm{cand}}$.
Since $\mathcal{C}_i$ contains the earlier states and selections, the choice can depend on the whole repair process so far.
Therefore, inspired by sequential recommendation~\cite{gru4rec, kang2018sasrec}, we formulate error type selection as a next-item prediction problem, where a trajectory corresponds to an interaction sequence, error types correspond to items, and \(\mathcal{T}^{(i)}_{\mathrm{cand}}\) denotes the set of candidate items available at round $i$.

The selector reads $\mathcal{C}_i$ as a sequence of $i$ positions, one for each state $(c_1,e_1),\ldots,(c_i,e_i)$.
Position $i$ pairs the $i$-th state with the error type $t_{i-1}$ selected in the round before it, where $t_0$ is a padding symbol.
Its input vector is encoded as follows:
\begin{equation}
    \mathbf{z}_j=
    \operatorname{LN}\!\left(
      \mathbf{W}_{s}[\operatorname{BGE}(c_j);
      \operatorname{BGE}(e_j)]\right)
    +\operatorname{LN}\!\left(
      \mathbf{W}_{t}\operatorname{Emb}_{t}(t_{j-1})\right)
    +\mathbf{p}_j,
    \label{eq:decision-embedding}
\end{equation}
where $\operatorname{BGE}(\cdot)$ is a pretrained text embedding model that produces frozen representations of the complete program and verifier feedback, $\operatorname{Emb}_{t}(\cdot)$ is a trainable error-type embedding initialized by applying $\operatorname{BGE}(\cdot)$ to the corresponding error-type name, and $\mathbf{p}_j$ is a positional embedding.
The matrices $\mathbf{W}_s$ and $\mathbf{W}_t$ are learnable projection matrices that map the representations into the same dimensional space, $[\,\cdot\,;\,\cdot\,]$ denotes concatenation, and $\operatorname{LN}(\cdot)$ denotes layer normalization.
The program and feedback channels distinguish states that report the same error types, which can favor different choices (Section~\ref{sec:motivation-study}).
Note that the error-type channel records which types the pipeline has already targeted and in which order.
The sequence is right-padded to a window length of $L=5$ to match the repair budget.

Subsequently, to model the decision dependencies among states at different rounds in the sequence, we employ a self-attention encoder~\cite{vaswani2017attention} to aggregate the sequence into a contextual vector:
\begin{equation}
    \mathbf{s}_i
    =\operatorname{AttEnc}_{\Theta}
    ([\mathbf{z}_1,\ldots,\mathbf{z}_i];
    \mathbf{M}_{\mathrm{causal}},\mathbf{M}_{\mathrm{pad}}),
    \label{eq:sequential-encoder}
\end{equation}
where $\operatorname{AttEnc}_{\Theta}(\cdot \, ; \, \cdot, \cdot)$ denotes the self-attention encoder network, \(\Theta\) denotes its parameters, and \(\mathbf{s}_i\) denotes the current decision context representation.
The causal mask $\mathbf{M}_{\mathrm{causal}}$ restricts each position to itself and earlier positions, as at actual deployment.
The padding mask $\mathbf{M}_{\mathrm{pad}}$ excludes the positions after a trajectory ends.

After obtaining \(\mathbf{s}_i\), we further measure its compatibility with each error type to identify the one best suited to the current decision state. Specifically, each error type is first encoded as an embedding in the same representation space and then scored against \(\mathbf{s}_i\).
The highest-scoring error type is selected as the one to be repaired in the current state, with the selection restricted to the candidate set. This process is formalized as follows:
\begin{equation}
    \begin{gathered}
        \hat{f}(t\mid\mathcal{C}_i)
        =\mathbf{s}_i^{\top}\mathbf{v}_t,
        \qquad
        \mathbf{v}_t
        =\operatorname{LN}(\mathbf{W}_t\operatorname{Emb}_t(t)),\\
        t_i
        =\operatorname*{arg\,max}_{
          t\in\mathcal{T}^{(i)}_{\mathrm{cand}}}
          \hat{f}(t\mid\mathcal{C}_i).
    \end{gathered}
    \label{eq:method-selection}
\end{equation}
This process implements the requirement of Section~\ref{sec:selection-problem} that $\pi$ choose among the types that the current feedback reports.
Notably, the vocabulary $\mathcal{T}$ combines the error types defined by our AutoVerus and VeruSAGE data-collection implementations, yielding the 35 types listed in Table~\ref{tab:error-type-vocabulary}.
Superscripts mark the default priority ranks from Section~\ref{sec:background-pipelines}, and rank 5 applies only to VeruSAGE.
\begin{table}[t]
    \centering
    \footnotesize
    \newcommand{\rk}[1]{\textsuperscript{\normalfont\,#1}}
    \caption{VeruSteer's vocabulary of 35 error types.}
    \label{tab:error-type-vocabulary}
    \begin{tabular}{@{}l>{\raggedright\arraybackslash\ttfamily}p{0.7\textwidth}@{}}
        \toprule
        \textbf{Category} & \normalfont\textbf{Error types} \\
        \midrule
        \multicolumn{2}{@{}l}{\textit{Rejected before verification}} \\
        \quad Rust syntax, types, and names
            & MismatchedType\rk{1}, TypeAnnotation, MethodNotFound, MissingImport,
              MissImpl, ExpectedParentheses \\
        \quad Verus language rules
            & ExecinGhost, CannotCallFunc, UnxProofBlock, RustAssert,
              UnsupportedForBitVector, LoopNoDec, EnsurePrivate, RequirePrivate,
              PubSpecVisibility, RequiresOldSelf \\
        \addlinespace
        \multicolumn{2}{@{}l}{\textit{Failed proof obligations}} \\
        \quad Contracts
            & PreCondFail, PreCondFailVecLen\rk{2}, PostCondFail, RecommendNotMet \\
        \quad Loops
            & InvFailFront\rk{4}, InvFailEnd\rk{5}, DecFailEnd, DecFailCont \\
        \quad Arithmetic
            & ArithmeticFlow\rk{3} \\
        \quad Assertions
            & AssertFail, BitVAssertFail, SplitAssertFail, SplitPreFail,
              SplitPostFail, TestAssertFail \\
        \quad Termination
            & TerminationFail, CannotProveTermination \\
        \quad Type invariants
            & ConstructorFailTypeInvariant \\
        \addlinespace
        \textit{Unrecognized message} & Other \\
        \bottomrule
    \end{tabular}
\end{table}

The error-type embedding table consists of embeddings for 35 error types and one padding embedding. Each type embedding is obtained by encoding its corresponding type name, and the configuration of the BGE encoder is provided in Section~\ref{sec:setup}.

\subsection{Dual-Loss Optimization and Deployment}
\label{sec:training-deployment}

As discussed in Section~\ref{sec:preference-supervision}, the selector learns from the two types of information contained in each paired sample $(\tau^+,\tau^-)\in\mathcal{D}_{\mathrm{pair}}$. 
Specifically, we optimize the selector by combining a cross-entropy loss with an auxiliary ranking loss. The cross-entropy loss supervises the recorded action at each valid step of $\tau^+$, whereas the ranking loss incorporates $\tau^-$ as an explicit comparison, providing the selector with supervision on relative repair outcomes in addition to action imitation.

First, for the cross-entropy loss term, the target at position \(i\) in \(\tau^+\) is the error type \(t_i^+\) selected at that position. 
The stepwise objective applies masked cross-entropy at every valid position:
\begin{equation}
    \mathcal{L}_{\mathrm{CE}}
    =-\mathbb{E}_{\mathcal{D}_{\mathrm{pair}}}\!\left[
      \frac{1}{L}\sum_{i=1}^{L}m_i
      \log P\bigl(t_i^+\mid\mathcal{C}_i^+\bigr)
    \right],
    \label{eq:ce-loss}
\end{equation}
where $\mathbb{E}_{\mathcal{D}_{\mathrm{pair}}}$ denotes
the average over training pairs and is estimated using mini-batch averages during training, $P(t\mid\mathcal{C}_i^+)$ is the softmax probability distribution computed from the scores $\hat{f}(\cdot\mid\mathcal{C}_i^+)$ restricted to $\mathcal{T}^{+(i)}_{\mathrm{cand}}$, and $m_i$ masks padding positions and positions randomly dropped during training.
In our implementation, we use the mask $m$ to randomly drop 10\% of otherwise valid positions as a form of regularization to improve generalization.

Next, for the auxiliary ranking loss, we use the selector's score of a trajectory's final selected error type under its final decision context as a surrogate trajectory score, $g(\tau) = \hat{f}(t_{\mathrm{last}}\mid \mathcal{C}_{\mathrm{last}})$.
The final decision context includes the preceding states and selections within the input window. Each sampled trajectory pair contributes one ranking term, and we do not construct additional ranking examples from its prefixes. Using the observed outcome preference $\tau^+\succ\tau^-$, the BPR objective~\cite{rendle2009bpr} encourages $g(\tau^+)>g(\tau^-)$:
\begin{equation}
\mathcal{L}_{\mathrm{BPR}}
=
-\mathbb{E}_{\mathcal{D}_{\mathrm{pair}}}
\left[
\log\sigma\bigl(g(\tau^+)-g(\tau^-)\bigr)
\right].
\label{eq:bpr-loss}
\end{equation}
where $\sigma$ denotes the sigmoid function.
Through the self-attention mechanism, this supervision updates the shared parameters that encode repair histories and score candidate actions. It provides an auxiliary preference
signal rather than directly optimizing trajectory selection probabilities.

Finally, by combining Equations~\eqref{eq:ce-loss} and ~\eqref{eq:bpr-loss}, the overall optimization objective is defined as:
\begin{equation}
    \min_{\Theta, \; \mathbf{w}} \mathcal{L} (\mathcal{D}_{\mathrm{pair}}),
    \qquad
    \mathcal{L}
    =\mathcal{L}_{\mathrm{CE}}
    +\lambda\mathcal{L}_{\mathrm{BPR}},
    \qquad \lambda=0.2.
    \label{eq:dual-loss}
\end{equation}
where $\Theta$ denotes the parameters of the self-attention encoder, and $\mathbf{w}$ denotes the remaining learnable parameters, including $\mathbf{W}_s$, $\mathbf{W}_t$, and $\mathbf{p}_i$.
In summary, we train a selector using preference pairs collected from the two pipelines.

At deployment, the pipeline runs as in Section~\ref{sec:repair-interaction}, with the selector in place of the default rule.
At round $i$, VeruSteer builds $\mathcal{C}_i$ from the run so far, selects $t_i$ with Equation~\ref{eq:method-selection}, and the pipeline executes $c_{i+1}=\operatorname{Repair}(c_i,e_i,t_i)$ before the next round.
The agents, acceptance checks, and verifier stay unchanged, and the run follows a single trajectory under the original budget.
During online execution, VeruSteer neither expands the repair tree nor introduces additional repair or verifier calls.
Algorithm~\ref{alg:verusteer} summarizes the training and deployment procedures.


\begin{algorithm}[t]
\small
\caption{Training and deployment of VeruSteer}
\label{alg:verusteer}
\begin{algorithmic}[1]
\REQUIRE Training, validation, and test instances;
repair pipeline; budget $B$
\STATE $\mathcal{D}_{\mathrm{pair}}\leftarrow\emptyset$
\FOR{each initial program $c_1$ of a training instance}
    \STATE $\mathcal{G}\leftarrow\textsc{ExpandTree}(c_1,B)$
    (Section~\ref{sec:tree-exploration})
    \STATE Add the pairs extracted from $\mathcal{G}$ to
    $\mathcal{D}_{\mathrm{pair}}$
    (Section~\ref{sec:preference-supervision})
\ENDFOR
\FOR{each hyperparameter configuration}
    \STATE Initialize $\Theta$ and $\mathbf{w}$
    \STATE Optimize $\Theta$ and $\mathbf{w}$ on
    $\mathcal{D}_{\mathrm{pair}}$ with
    Equation~\eqref{eq:dual-loss}, using validation
    score $G$ for early stopping and checkpoint selection
\ENDFOR
\STATE Retain the checkpoint with the highest validation
score $G$ across configurations
\STATE \textbf{Deployment and evaluation:}
\FOR{each test instance}
    \FOR{each round $i$ until the run stops or budget $B$ is exhausted}
        \STATE Build $\mathcal{C}_i$ and
        $\mathcal{T}^{(i)}_{\mathrm{cand}}$
        \STATE Select $t_i$ with
        Equation~\eqref{eq:method-selection}
        \STATE $c_{i+1}\leftarrow
        \operatorname{Repair}(c_i,e_i,t_i)$
    \ENDFOR
    \STATE Record the resulting trajectory for evaluation
\ENDFOR
\end{algorithmic}
\end{algorithm}
At each decision, inference computes two frozen text embeddings for the current program and verifier feedback, followed by $O(L^2d)$ self-attention and $O(|\mathcal{T}^{(i)}_{\mathrm{cand}}|d)$ candidate scoring for hidden dimension $d$.

\section{Evaluation}
\label{sec:experiments}

Our evaluation addresses the following research questions:

\begin{enumerate}[label=\textbf{RQ\arabic*.}, leftmargin=3.4em, labelsep=0.6em, itemsep=2pt, topsep=4pt]
    \item \textit{Overall performance.} How well does VeruSteer preserve and improve the default rule's repair outcomes?
    \item \textit{Ablation study.} How do the error-type selection method, the training losses and input representations, and the input sequence length of VeruSteer affect its repair outcomes?
    \item \textit{Repair progress across rounds.} How do the numbers of verified proof obligations and reported errors change over the repair rounds under VeruSteer and under the default rule?
    \item \textit{Choices among reported error types.} When a state reports several error types, how do the choices of VeruSteer and their verification outcomes differ from those of the default rule?
\end{enumerate}

We first describe the pipelines, data, training, and evaluation procedure (Section~\ref{sec:setup}).
We then examine repair outcomes in RQ1, ablate the selector in RQ2, and study repair progress and individual choices in RQ3 and RQ4.

\subsection{Experimental Setup}
\label{sec:setup}
\noindent\textbf{Repair pipelines and LLMs.}
We use AutoVerus~\cite{AutoVerus} and VeruSAGE~\cite{VeruSAGE} as the underlying repair pipelines.
Both select one error type per round and dispatch the repair agent specialized for it (Section~\ref{sec:background-pipelines}).
VeruSteer is called at this dispatch point and ranks the types extracted from the current verifier output, with the same agents, acceptance checks, and verifier.
The initial programs are generated by DeepSeek-Chat or Qwen-Coder (i.e., Qwen3-Coder-480B-A35B-Instruct), each under zero-shot and five-shot prompting, which gives four generation configurations per pipeline.
All repair rounds use DeepSeek-Chat.

\noindent\textbf{Datasets and splits.}
The StarVerus benchmark contains 762 verification tasks~\cite{jiang2026starverus}.
A task is one benchmark problem, independent of the generation model, prompting setting, and repair pipeline.
Each generation configuration produces at most one initial program for a task, so a task has up to four initial programs.
Each initial program is repaired by both pipelines.
An initial program that already passes Verus has nothing to repair and yields no repair tree.
684 of the 762 tasks have at least one repair tree and form our corpus; the other 78 have none.

Each repair tree is built from one initial program and one repair pipeline.
It explores alternative error-type selections from the initial state (Section~\ref{sec:tree-exploration}), and each root-to-leaf path is a repair trajectory.
In total, 1,614 initial programs have repair trees, and they yield 3{,}224 trees: 1,613 for AutoVerus and 1,611 for VeruSAGE.
Every tree contains the default trajectory, and 718 of the 3{,}224 trees also contain alternative trajectories.

We split the 684 tasks into 342 training, 69 validation, and 273 test tasks, approximately 5:1:4.
The split is by task: all trees of a task fall in the same split, so no task contributes to both training and evaluation.
Table~\ref{tab:corpus_statistics} reports the size of each split.
The training trees provide 18{,}885 preference pairs (Section~\ref{sec:preference-supervision}) for training the selector, and the validation trees are used to select its checkpoint.

\begin{table}[t]
  \centering
  \small
  \caption{Size of each split. Trees with alternatives contain at least one trajectory besides the default one.}
  \label{tab:corpus_statistics}
  \begin{tabular*}{\linewidth}{@{\extracolsep{\fill}}lrrrr@{}}
    \toprule
    & \textbf{Train} & \textbf{Validation} & \textbf{Test} & \textbf{All} \\
    \midrule
    Tasks & 342 & 69 & 273 & 684 \\
    \midrule
    Repair trees & 1,636 & 318 & 1,270 & 3,224 \\
    \quad AutoVerus & 819 & 159 & 635 & 1,613 \\
    \quad VeruSAGE & 817 & 159 & 635 & 1,611 \\
    \midrule
    Trees with alternatives & 359 & 70 & 289 & 718 \\
    \quad AutoVerus & 198 & 40 & 159 & 397 \\
    \quad VeruSAGE & 161 & 30 & 130 & 321 \\
    \midrule
    Repair trajectories & 36,028 & 3,044 & 17,356 & 56,428 \\
    \quad AutoVerus & 20,014 & 1,341 & 10,104 & 31,459 \\
    \quad VeruSAGE & 16,014 & 1,703 & 7,252 & 24,969 \\
    \bottomrule
  \end{tabular*}
  \vspace{-10pt}
\end{table}

\noindent\textbf{Implementation and training.}
For VeruSteer, \(\operatorname{BGE}(\cdot)\) employs the pretrained BGE-M3\footnote{https://huggingface.co/BAAI/bge-m3} embedding model~\cite{chen2024m3embedding}, which separately encodes the program, verifier feedback, and error type associated with each state into 1,024-dimensional vectors. These representations are then projected into a 64-dimensional space through the projection matrices \(\mathbf{W}_s\) and \(\mathbf{W}_t\). The self-attention encoder \(\operatorname{AttEnc}_{\Theta}(\cdot)\) consists of a single-head self-attention module followed by a position-wise feed-forward layer, with a dropout rate of 0.1. The complete selector takes an input sequence of length \(L=5\), covering the latest five states.
We train the selector on preference-pair data collected from the two pipelines using the objective in Equation~\eqref{eq:dual-loss}. The selector is optimized using Adam. The learning rate is selected from \(\{10^{-3}, 5\times10^{-4}, 10^{-4}, 5\times10^{-5}, 10^{-5}\}\), and the weight decay is selected from \(\{10^{-3}, 10^{-4}, 10^{-5}, 10^{-6}\}\). The batch size is set to 256. 
For early stopping and checkpoint selection, we use the validation score $G=\mathrm{EM}_v+\tfrac12(\mathrm{EM}_s+\mathrm{EM}_h)$,
where $\mathrm{EM}_v$, $\mathrm{EM}_s$, and $\mathrm{EM}_h$ are the sequence exact-match rates on the validation set
for the verification, shorter-path, and higher-score categories, respectively.
A trajectory is an exact match if, given its recorded history prefixes, the selector predicts the recorded action from the legal candidates at every valid step. The verification category receives twice the weight of either other category to reflect its greater importance. Training stops if $G$ does not improve for ten consecutive epochs or after 50 epochs, and the checkpoint with the highest $G$ is retained.
Each pipeline is evaluated on the test set using its corresponding checkpoint selected on the validation set.
\noindent\textbf{Evaluation protocol.}
The default rule follows the preset priority over error types of each pipeline (Section~\ref{sec:background-pipelines}).
We evaluate a selector by replaying its choices on the stored test trees of the corresponding pipeline.
Each tree already records the outcome of repairing every reported type at each state (Section~\ref{sec:tree-exploration}), so replay needs no new LLM or verifier calls.
Starting at the root, the selector picks one of the types reported at the current state, and the replay moves to the stored child state for that type.
The replay stops at a leaf, that is, when the program verifies, the feedback yields no candidate type, or the budget of $B=5$ rounds is spent.
The path from the root to that leaf is the selector's trajectory.
Replaying the default rule returns the stored default trajectory of every test tree.

\noindent\textbf{Evaluation metrics.}
On each test tree, we compare the selector's trajectory with the default trajectory under the three cases of Equation~\ref{eq:trajectory-preference}, each of which gives one objective.
\textit{Verification} asks whether the final program verifies, that is, whether the stored Verus output for the final state reports zero errors.
When both trajectories verify, \textit{Shorter path} compares their numbers of repair rounds.
When both fail, \textit{Higher score} compares their final progress scores $Q$ (Section~\ref{sec:trajectory-quality}).

We report each objective on the trees with alternative trajectories, split into \textit{Retention} and \textit{Exploration}.
Retention contains the trees where the default trajectory is already optimal under the objective, and a selector succeeds if its trajectory is at least as good.
Exploration contains the trees with a strictly better alternative trajectory, and a selector succeeds if it improves on the default outcome.
The success rate is the share of these trees on which the selector succeeds.
RQ1 defines the trees of each group for each objective and reports both pipelines, and RQ2 ablates VeruSteer on the Verification groups of VeruSAGE.

\subsection{RQ1: Overall Performance}
RQ1 examines whether VeruSteer preserves the outcomes already achieved by the default rule and finds better alternatives when they are available.
Table~\ref{tab:overall_verification} first reports the overall verification benefit on all test trees.
AutoVerus improves from 96 to 98 verified trees out of 635, gaining three and losing one, for a net increase of two trees (0.31 percentage points).
VeruSAGE improves from 89 to 96 verified trees out of 635, gaining nine and losing two, for a net increase of seven trees (1.10 percentage points).
Pooled across both pipelines, this is 185 to 194 verified trees out of 1,270: 12 gained, three lost, and a net increase of nine trees (0.71 percentage points).

\begin{table}[t]
  \centering
  \small
  \setlength{\tabcolsep}{3pt}
  \caption{Verification on the test trees. Default rule and VeruSteer report verified trees and verification rates; gains and losses are relative to the default rule.}
  \label{tab:overall_verification}
  \begin{tabular*}{\linewidth}{@{\extracolsep{\fill}}lrrrrrrr@{}}
    \toprule
    Baseline & Test trees & Default rule & VeruSteer & Gained & Lost & Net & $\Delta$ (pp) \\
    \midrule
    AutoVerus & 635 & 96 (15.12\%) & 98 (15.43\%) & 3 & 1 & \textbf{+2} & \textbf{+0.31} \\
    VeruSAGE & 635 & 89 (14.02\%) & 96 (15.12\%) & 9 & 2 & \textbf{+7} & \textbf{+1.10} \\
    Total & 1270 & 185 (14.57\%) & 194 (15.28\%) & 12 & 3 & \textbf{+9} & \textbf{+0.71} \\
    \bottomrule
  \end{tabular*}
  \vspace{-10pt}
\end{table}

Table~\ref{tab:main_results} then reports the three objectives on the test trees with recorded alternative trajectories, 159 for AutoVerus and 130 for VeruSAGE (Table~\ref{tab:corpus_statistics}).
Each objective defines its eligible trees separately.
For Verification, Retention contains trees where the default trajectory verifies, and Exploration contains trees where the default trajectory fails and at least one alternative trajectory verifies.
For Shorter path, the default trajectory and at least one alternative trajectory verify.
Its Retention group contains trees where the default needs no more rounds than any verified alternative, and its Exploration group contains trees with a shorter verified alternative.
These trees also belong to Verification Retention.
For Higher score, all trajectories fail, and the groups are determined by comparing their final progress scores.

Each cell reports successful trees divided by eligible trees, followed by the success rate.
The Retention rate is the share of trees where the selector preserves the default-optimal outcome, and the Exploration rate is the share where it improves on the default.
Empty groups are marked with a dash and $n=0$; Total pools counts across the four configurations.

\begin{table}[t]
  \centering
  \small
  \setlength{\tabcolsep}{3pt}
  \caption{Decision quality on test trees with default and alternative trajectories. Each cell reports successful / eligible trees and the success rate.}
  \label{tab:main_results}
  \begin{tabular*}{\linewidth}{@{\extracolsep{\fill}}llccccc@{}}
    \toprule
    \multirow{2}{*}{Baseline} & \multirow{2}{*}{Objective} &
    \multicolumn{2}{c}{DeepSeek-Chat} & \multicolumn{2}{c}{Qwen-Coder} & \multirow{2}{*}{Total} \\
    \cmidrule(lr){3-4}\cmidrule(lr){5-6}
    & & 0-shot & 5-shot & 0-shot & 5-shot & \\
    \midrule
    \multicolumn{7}{l}{\textbf{Retention (default-optimal): preserve the default-optimal outcome}} \\
    \midrule
    \multirow{3}{*}{AutoVerus} & Verification & --- ($n=0$) & 5/6 (83.3\%) & 3/3 (100.0\%) & 6/6 (100.0\%) & 14/15 (93.3\%) \\
     & Shorter path & --- ($n=0$) & 2/3 (66.7\%) & 2/2 (100.0\%) & 1/1 (100.0\%) & 5/6 (83.3\%) \\
     & Higher score & 33/33 (100.0\%) & 19/20 (95.0\%) & 32/33 (97.0\%) & 16/16 (100.0\%) & 100/102 (98.0\%) \\
    \cmidrule(lr){1-7}
    \multirow{3}{*}{VeruSAGE} & Verification & 1/1 (100.0\%) & 3/5 (60.0\%) & 3/3 (100.0\%) & 5/5 (100.0\%) & 12/14 (85.7\%) \\
     & Shorter path & 1/1 (100.0\%) & 1/2 (50.0\%) & 1/2 (50.0\%) & 4/5 (80.0\%) & 7/10 (70.0\%) \\
     & Higher score & 26/27 (96.3\%) & 12/14 (85.7\%) & 24/24 (100.0\%) & 14/14 (100.0\%) & 76/79 (96.2\%) \\
    \midrule
    \multicolumn{7}{l}{\textbf{Exploration (default-suboptimal): select an improving alternative trajectory}} \\
    \midrule
    \multirow{3}{*}{AutoVerus} & Verification & 1/1 (100.0\%) & 0/1 (0.0\%) & 1/1 (100.0\%) & 1/1 (100.0\%) & 3/4 (75.0\%) \\
     & Shorter path & --- ($n=0$) & 1/3 (33.3\%) & 0/1 (0.0\%) & 1/4 (25.0\%) & 2/8 (25.0\%) \\
     & Higher score & 0/3 (0.0\%) & 4/9 (44.4\%) & 4/14 (28.6\%) & 1/12 (8.3\%) & 9/38 (23.7\%) \\
    \cmidrule(lr){1-7}
    \multirow{3}{*}{VeruSAGE} & Verification & 2/2 (100.0\%) & 3/4 (75.0\%) & 3/3 (100.0\%) & 1/2 (50.0\%) & 9/11 (81.8\%) \\
     & Shorter path & --- ($n=0$) & 2/2 (100.0\%) & 1/1 (100.0\%) & --- ($n=0$) & 3/3 (100.0\%) \\
     & Higher score & 2/4 (50.0\%) & 4/7 (57.1\%) & 3/7 (42.9\%) & 1/8 (12.5\%) & 10/26 (38.5\%) \\
    \bottomrule
  \end{tabular*}
  \vspace{-10pt}
\end{table}

For Verification, the Retention rates are 93.3\% on AutoVerus and 85.7\% on VeruSAGE, and the Exploration rates are 75.0\% and 81.8\%.
All the drops and missed improvements fall in the five-shot configurations, while all nonempty zero-shot Verification groups show full retention or full recovery.

For Shorter path, the Retention rates are 83.3\% on AutoVerus and 70.0\% on VeruSAGE, and the Exploration rates are 25.0\% (2/8) and 100.0\% (3/3).
The improvement in repair rounds is therefore more consistent on VeruSAGE, while both pipelines lose some default shorter paths.

For Higher score, the Retention rates are 98.0\% and 96.2\%, and the Exploration rates are 23.7\% and 38.5\%.

\noindent\textbf{Takeaway.}
VeruSteer yields nine net additional verified test trees across the two pipelines.
On trees with recorded alternatives, it preserves most default outcomes and recovers 3/4 and 9/11 of the available verification opportunities on AutoVerus and VeruSAGE, respectively.
It also finds shorter successful trajectories and higher final scores on both pipelines where such alternatives exist.

\subsection{RQ2: Ablation Study}
\label{sec:rq2}
\begin{figure}[t]
    \centering
    \includegraphics[width=\linewidth]{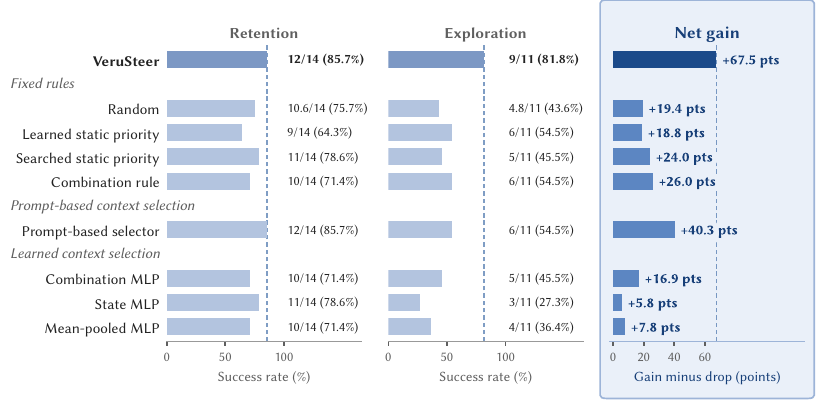}
    \caption{Ablation on the error-type selection method.}
    \Description{Bars for VeruSteer and for eight other selectors in three groups. Fixed rules: Random, learned static priority, searched static priority, and the combination rule. Prompt-based context selection: the prompt-based selector. Learned context selection: VeruSteer, Combination MLP, State MLP, and Mean-pooled MLP. State MLP reads only the current state, whereas Mean-pooled MLP averages the valid historical steps. Three horizontal bar charts share their rows. The left and middle charts show the Retention and Exploration success rates from 0 to 100 percent, each labeled with the numbers of successful and eligible trees and the rate, and the right chart shows the net gain, the Exploration success rate minus the Retention drop from 100 percent, in percentage points, on the Verification objective. A dashed vertical line marks the value of VeruSteer, drawn in dark blue.}
    \label{fig:ablation_selection}
\end{figure}

\begin{figure}[t]
    \centering
    \includegraphics[width=\linewidth]{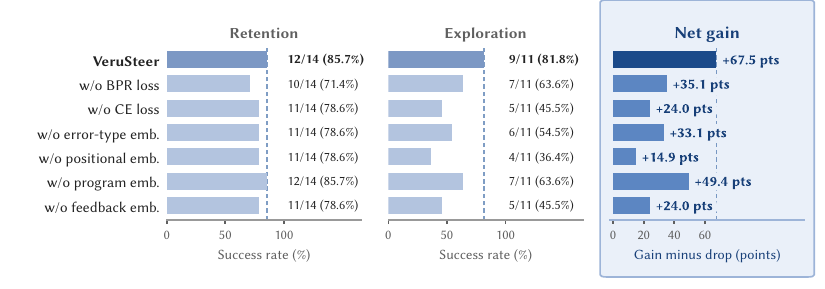}
    \caption{Ablation on the training losses and input representations for VeruSteer.}
    \Description{Bars for VeruSteer and for the variants without the BPR loss, the CE loss, the error-type embedding, the positional embedding, the program embedding, or the feedback embedding. Three horizontal bar charts share their rows. The left and middle charts show the Retention and Exploration success rates from 0 to 100 percent, each labeled with the numbers of successful and eligible trees and the rate, and the right chart shows the net gain, the Exploration success rate minus the Retention drop from 100 percent, in percentage points, on the Verification objective. A dashed vertical line marks the value of VeruSteer, drawn in dark blue.}
    \label{fig:ablation_components}
\end{figure}

\begin{figure}[t]
    \centering
    \includegraphics[width=\linewidth]{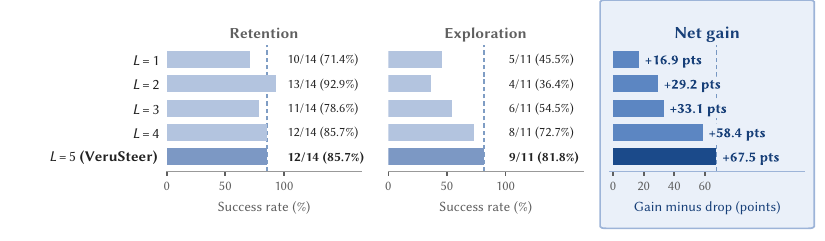}
    \caption{Ablation on the input sequence length for VeruSteer.}
    \Description{Bars for selectors that read the latest 1, 2, 3, or 4 states and for VeruSteer, which reads 5. Three horizontal bar charts share their rows. The left and middle charts show the Retention and Exploration success rates from 0 to 100 percent, each labeled with the numbers of successful and eligible trees and the rate, and the right chart shows the net gain, the Exploration success rate minus the Retention drop from 100 percent, in percentage points, on the Verification objective. A dashed vertical line marks the value of VeruSteer, drawn in dark blue.}
    \label{fig:ablation_length}
\end{figure}

RQ2 asks which parts of VeruSteer account for its gains.
We run three ablations on VeruSAGE: the error-type selection method, the training losses and input representations, and the input sequence length.
Figures~\ref{fig:ablation_selection}--\ref{fig:ablation_length} report the Verification objective of RQ1, pooled over the four generation configurations.
The left and middle charts show the Retention and Exploration rates with the numbers of successful and eligible trees, as in Table~\ref{tab:main_results}.
The right chart shows the net gain, the Exploration rate minus the Retention drop (100\% minus the Retention rate), in percentage points; a larger value means a better trade-off.

\noindent\textbf{Ablation on the error-type selection method.}
Since the best choice for the same reported types varies across states (Section~\ref{sec:motivation-study}), VeruSteer reads the program, the verifier feedback, and the earlier selections.
This ablation asks how a selector should use such context, and compares VeruSteer with three kinds of selectors that all choose among the reported types:
\begin{itemize}[leftmargin=1.6em, labelsep=0.5em, itemsep=2pt, topsep=2pt]
    \item \textbf{Fixed rules} choose from the reported types alone, ignoring the program and the feedback.
    \begin{itemize}[leftmargin=1.4em, labelsep=0.4em, itemsep=0pt, topsep=1pt]
        \item \textbf{Random} chooses uniformly at random; we report the mean over five seeds.
        \item \textbf{Learned static priority} ranks the types by one global score, the share of pairwise comparisons on the training trees that a type wins under Equation~\ref{eq:trajectory-preference}.
        \item \textbf{Searched static priority} ranks the types by one global order, searched directly for the most verified training trees under replay.
        \item \textbf{Combination rule} learns the scores of the learned static priority separately for each set of types reported together.
    \end{itemize}
    \item \textbf{Prompt-based context selection} asks an LLM to choose the type.
    \begin{itemize}[leftmargin=1.4em, labelsep=0.4em, itemsep=0pt, topsep=1pt]
        \item \textbf{Prompt-based selector} uses DeepSeek-Flash (temperature 0.3, thinking disabled) with one answer per decision. Its prompt includes the current program, verifier feedback, reported types, repair history, and examples retrieved from training trees.
    \end{itemize}
    \item \textbf{Learned context selection} trains the selector on the training trees, as VeruSteer does, but with a different encoder.
    \begin{itemize}[leftmargin=1.4em, labelsep=0.4em, itemsep=0pt, topsep=1pt]
        \item \textbf{Combination MLP} replaces the self-attention encoder with an MLP over the set of reported types, trained with the cross-entropy loss only.
        \item \textbf{State MLP} replaces it with an MLP over the program and feedback of the current state, trained with both losses.
        \item \textbf{Mean-pooled MLP} replaces the selector with an MLP applied to the mean-pooled state representations from the latest $L=5$ steps, trained with both losses.
    \end{itemize}
\end{itemize}

Figure~\ref{fig:ablation_selection} shows that VeruSteer outperforms all eight alternatives.
It keeps 12 of the 14 default successes (85.7\%), which no alternative exceeds, and recovers 9 of the 11 available verification opportunities (81.8\%), against at most 6 (54.5\%) for the alternatives.
Its net gain of 67.5 points is 1.7 times that of the best alternative, the prompt-based selector (40.3 points).

The fixed rules commit to one type for every state that reports the same types, although the best type for such states differs, and fitting the order directly to the training trajectories does not help: the Searched static priority recovers 5/11 opportunities (45.5\%), fewer than the Learned static priority with 6/11 (54.5\%).
The prompt-based selector reads the context and is the strongest alternative: it keeps 12/14 default successes, as VeruSteer does, and recovers 6/11 opportunities (54.5\%), so its net gain of 40.3 points is well above those of the fixed rules (18.8--26.0) and the learned MLPs (5.8--16.9).
It still recovers fewer opportunities than VeruSteer (6/11 against 9/11), which is trained on all the trajectory comparisons instead of reading a few retrieved examples.

Among the learned selectors, the MLPs separate whether the selector can see the history from how it encodes it.
State MLP sees only the current state; it keeps 11/14 default successes (78.6\%) and recovers 3/11 opportunities (27.3\%).
Mean-pooled MLP also sees the history and recovers 4/11 (36.4\%), but it keeps only 10/14 (71.4\%), so its net gain is 7.8 points, against 5.8 for State MLP.
Seeing the history thus brings little when it is simply averaged, whereas VeruSteer, whose self-attention encoder reads the same window in order, reaches 67.5 points.

\noindent\textbf{Ablation on the training losses and input representations for VeruSteer.}
VeruSteer combines two training losses with several representations of a state.
This ablation asks whether each is needed and which goal it serves, by removing them one at a time and training each variant on the same preference pairs:
\begin{itemize}[leftmargin=1.6em, labelsep=0.5em, itemsep=1pt, topsep=2pt]
    \item \textbf{w/o BPR loss} drops the pairwise ranking loss, which scores the preferred trajectory above the other one in a pair (Section~\ref{sec:training-deployment}).
    \item \textbf{w/o CE loss} drops the cross-entropy loss, which imitates the selections along the preferred trajectory step by step.
    \item \textbf{w/o error-type emb.} drops the embedding of the error types the pipeline has already targeted.
    \item \textbf{w/o positional emb.} drops the embedding of the position of a state in the sequence.
    \item \textbf{w/o program emb.} drops the embedding of the program.
    \item \textbf{w/o feedback emb.} drops the embedding of the verifier feedback.
\end{itemize}

Figure~\ref{fig:ablation_components} shows that every removal weakens VeruSteer, and in different ways.
Step-wise imitation fits the type chosen at each step of the preferred trajectory; without the CE loss, Exploration falls from 9/11 to 5/11 (45.5\%) and Retention from 12/14 to 11/14 (78.6\%).
Pairwise ranking compares the outcomes of two trajectories; without the BPR loss, Retention falls from 12/14 to 10/14 (71.4\%), while Exploration remains 7/11 (63.6\%).
Removing the CE loss thus costs more Exploration (9/11 to 5/11) than removing the BPR loss (to 7/11), and removing the BPR loss costs more Retention.
Step-wise imitation alone finds many better alternatives but cannot tell when leaving the default trajectory is not worth it, and the ranking over whole trajectories supplies that judgment.

Each representation contributes, since removing any of them lowers the net gain from 67.5 points.
The positional embedding, which tells the selector where a state lies in the sequence, matters most: without it the value falls to 14.9 points, as Exploration falls to 4/11 (36.4\%) and Retention to 11/14 (78.6\%).
Removing the feedback comes next, with 24.0 points (Exploration 5/11, Retention 11/14), followed by the error-type embedding, which tells the selector which types were already targeted, with 33.1 points (6/11 and 11/14).
The positional and error-type embeddings together encode the repair history, and removing either one cuts the value by half or more.
The program embedding matters least: without it the value is still 49.4 points (Exploration 7/11, Retention 12/14), probably because the feedback already names the failing types and where they occur, so the selector can recover part of the program from the remaining inputs.

\noindent\textbf{Ablation on the input sequence length for VeruSteer.}
VeruSteer reads a window of the latest $L=5$ positions of the trajectory, which matches the budget of $B=5$ rounds.
A shorter window leaves out the earlier states of the trajectory, so this ablation asks how long a window the selector needs.
We shorten the window to $L=1,2,3,4$.
Position $i$ pairs the $i$-th state with the error type selected in the round before it (Section~\ref{sec:selector}), so even $L=1$ sees the previous selection and lacks only the earlier states.

Figure~\ref{fig:ablation_length} shows that a longer window improves Exploration.
Exploration is 5/11, 4/11, 6/11, and 8/11 for $L=1$ to $4$, and 9/11 for VeruSteer, so that the net gain grows from 16.9 points at $L=1$ to 67.5 points at $L=5$.
Retention varies irregularly with $L$ (10/14, 13/14, 11/14, and 12/14 for $L=1$ to $4$, against 12/14 for VeruSteer).
$L=2$ has the highest Retention but the lowest Exploration, so it keeps most default outcomes at the cost of finding few better alternatives.
Retention thus depends on how often a selector departs from the default trajectory as well as on how well it chooses, and the net gain shows that a longer window improves the trade-off between the two.

\noindent\textbf{Takeaway.}
The gains of VeruSteer come from what its selector reads, and each part of it contributes.
All eight alternative selectors leave the net gain at 40.3 points or less, against 67.5 for VeruSteer; the best of them reads the context in a prompt, and retaining history through mean pooling reaches only 7.8.
Removing either loss or any input representation lowers it to between 14.9 and 49.4, and the two losses divide the work, with step-wise imitation serving Exploration and pairwise ranking serving Retention.
A longer window of earlier selections raises it steadily, from 16.9 points at $L=1$ to 67.5 at $L=5$.

\subsection{RQ3: Repair Progress across Rounds}
\begin{figure}[t]
    \centering
    \includegraphics[width=0.9\linewidth]{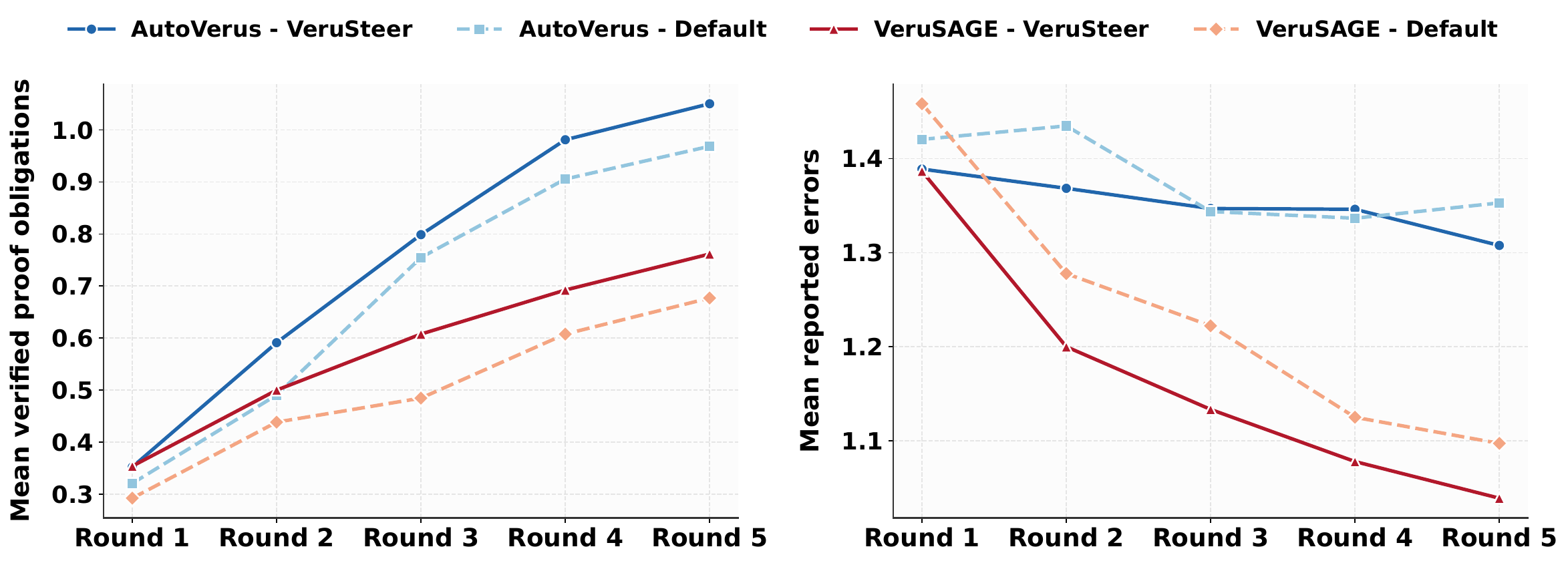}
    \caption{Mean verified proof obligations and reported errors per round.}
    \Description{Two line plots side by side over repair rounds 1 to 5. The left plot shows the mean number of verified proof obligations and the right plot the mean number of reported errors, each for the default rule and VeruSteer on AutoVerus and VeruSAGE.}
    \label{fig:avg_score}
    \vspace{-10pt}
\end{figure}

RQ3 asks whether the gains in RQ1 reflect steady progress during repair or only a difference in the final outcome.
Figure~\ref{fig:avg_score} tracks Rounds 1--5 on the test trees that contain both default and alternative trajectories (159 for AutoVerus, 130 for VeruSAGE).
It reports the mean number of verified proof obligations and the mean number of reported errors, the two components of the progress score $Q$ (Section~\ref{sec:trajectory-quality}), each for the default rule and VeruSteer on the same trees.
A trajectory that terminates early keeps its final state in the later rounds, and invalid negative error counts are excluded from the error averages.

VeruSteer has more verified proof obligations than the default rule from the first round onward and keeps the lead.
The gap is already 0.031 on AutoVerus and 0.062 on VeruSAGE at Round 1, and it reaches 0.082 and 0.085 at Round 5, so the final improvement (1.050 against 0.969 on AutoVerus, 0.762 against 0.677 on VeruSAGE) builds up over the rounds and does not appear only at the end.
The error count shows the same pattern.
On VeruSAGE, VeruSteer has fewer reported errors than the default at every round, by 0.05 to 0.09, and ends at 1.04 against 1.10.
On AutoVerus, it has fewer errors at Rounds 1, 2, and 5, ending at 1.31 against 1.35, and it differs from the default by less than 0.01 at Rounds 3 and 4.
On both pipelines, VeruSteer thus ends the repair with more verified proof obligations and fewer reported errors than the default, so the higher score is not bought with additional errors.

\noindent\textbf{Takeaway.}
The advantage of VeruSteer builds up steadily during repair and does not appear only in the final outcome.
On both pipelines, it leads the default rule in verified proof obligations from Round 1 and widens the lead to 0.082 on AutoVerus and 0.085 on VeruSAGE by Round 5, while also ending with fewer reported errors (1.31 against 1.35 and 1.04 against 1.10).

\subsection{RQ4: Choices among Reported Error Types}
\begin{figure}[t]
    \centering
    \includegraphics[width=1.0\linewidth]{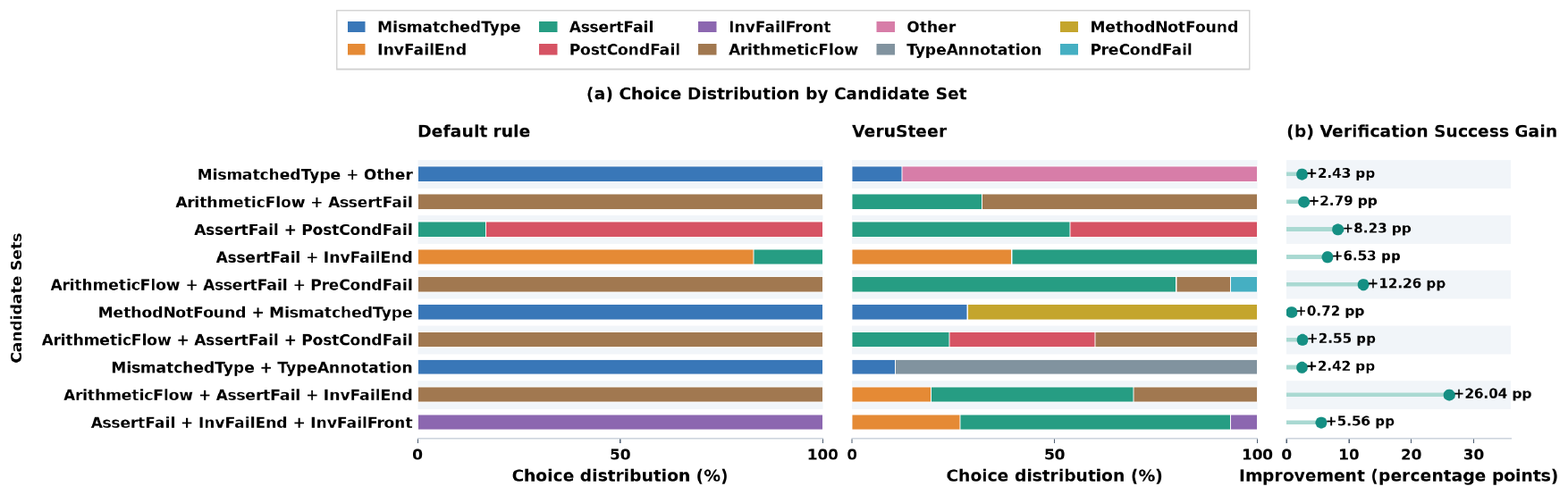}
    \caption{Choice distributions (a) and associated verification-rate differences (b) for the ten most frequent candidate sets with positive differences in the full corpus. Selector-specific state populations differ; pp denotes percentage points.}
    \Description{Stacked bars show the default rule's and VeruSteer's error-type choices; dots show VeruSteer minus default in associated final verification rates.}
    \label{fig:error_type_combination}
    \vspace{-10pt}
\end{figure}

Figure~\ref{fig:error_type_combination} groups multi-error states by their candidate set, pooling both pipelines and all generation configurations over the full corpus, including training tasks.
It shows the ten most frequent candidate sets with positive observed verification-rate differences, providing a descriptive view of policy behavior rather than a test of generalization.
Panel (a) reports each selector's choice distribution, counting each state once per tree; panel (b) reports the associated final verification-rate difference, VeruSteer minus default, in percentage points (pp).
The two selectors contribute different state populations, so these are descriptive comparisons rather than paired effects of changing a choice at the same state.

The default rule usually concentrates on one type, while VeruSteer often departs from that choice.
For \texttt{MismatchedType}+\texttt{Other}, the default always selects \texttt{MismatchedType}, whereas VeruSteer selects \texttt{Other} in 87.74\% of its decisions.
The associated verification rates are $2/3618$ and $17/685$, respectively, a difference of 2.43 pp.
The displayed differences range from 0.72 to 26.04 pp, with the largest for \texttt{ArithmeticFlow}+\texttt{AssertFail}+\texttt{InvFailEnd}.

\noindent\textbf{Takeaway.}
VeruSteer makes context-dependent choices among reported error types,
departing from the default rule's near-deterministic selections.
In the candidate sets shown, these choices are associated with
0.72--26.04 percentage-point higher final verification rates.
These observations illustrate how contextual selection can identify
repair opportunities beyond preset error-type priorities.

\section{Discussion and Threats to Validity}
\label{sec:discussion}

This section discusses what the results mean for verification automation, the limitations and directions that follow, and the threats to the validity of the evaluation.

\subsection{Implications for Verification Automation}
\label{sec:implications}

\noindent\textbf{Repair order is a separate, learnable part of verification automation.}
Work on LLM-based repair for Verus mostly strengthens what happens once a failure is targeted, through retrieved context, counterexamples, tree search, and candidate ranking (Sections~\ref{sec:background-pipelines} and~\ref{sec:related-repair}).
Which reported error type to repair next is a different axis, and improving it leaves the repair agents untouched: with the same agents, acceptance checks, verifier, and budget, VeruSteer recovers 75.0\% and 81.8\% of the available verification opportunities on AutoVerus and VeruSAGE (RQ1).
The repair trees show why this can be learned, since the best choice varies with the state even for the same reported types (Section~\ref{sec:motivation-study}).
A pipeline can therefore treat error-type selection as a policy learned from repair outcomes, instead of a priority list fixed when the pipeline is written.

\noindent\textbf{Casting error selection as sequential recommendation makes the repair history usable.}
VeruSteer reads a trajectory as an interaction sequence, error types as items, and the reported types of each round as the candidate set (Section~\ref{sec:selector}), so the choice of a repair target becomes next-item prediction conditioned on the decision context, which includes the earlier states and selections.
RQ2 shows that the borrowed components matter: a selector that reads only the current state recovers 3 of the 11 verification opportunities (27.3\%) against 9 of 11 (81.8\%) for VeruSteer, the net gain grows from 16.9 points with a window of one state to 67.5 points with the full window of $L=5$, and the pairwise ranking keeps the Retention drop at 14.3\% where its removal doubles it to 28.6\%.
Only the choice of which failure to target is learned here, while the other two decisions of a repair round, how to repair it and which candidate to carry forward, stay with the pipeline (Section~\ref{sec:background-pipelines}).
We expect the same casting to fit other LLM pipelines that dispatch an agent from a candidate set that changes every round, although we evaluate only the two Verus pipelines.

\noindent\textbf{Verifier outcomes supply supervision and a controlled evaluation without manual labels.}
Verus decides every outcome mechanically, so alternative trajectories from the same program can be compared directly by verification, repair rounds, and progress (Section~\ref{sec:trajectory-quality}).
Successful and failed trajectories alike then yield preferences, and the 18{,}885 preference pairs in the training partition need no manual annotation, with VeruSteer also improving final progress on trees where no recorded trajectory verifies (RQ1).
The stored trees can be replayed without new LLM or verifier calls, so every selector is compared on the same recorded transitions (Section~\ref{sec:setup}).
Reporting Retention and Exploration separately keeps this comparison informative, since a selector can look competitive in one and lose in the other: the learned static priority recovers 6 of the 11 verification opportunities (54.5\%), the most that any of the eight alternative selectors recovers, yet keeps only 9 of the 14 default successes (64.3\%), fewer than any of them (RQ2).

\subsection{Limitations and Future Directions}
\label{sec:limitations}

\noindent\textbf{The selector learns only which failure to repair.}
A repair round makes three decisions: which failure to target, how to repair it, and which candidate to carry forward (Section~\ref{sec:background-pipelines}), and VeruSteer learns only the first.
It acts only at states that report several error types, since a single reported type leaves nothing to select.
The other two decisions stay inside the repair agent, which picks a repair method and generates several candidate patches, and our trees keep only the patch it accepts, so they do not show what a different method or a different candidate would have led to.
Learning these decisions would need trees that keep such alternatives.

\noindent\textbf{Supervision comes from a finite and costly exploration.}
At each state, a tree samples the repair of each reported error type once and keeps the accepted result as the successor, so another LLM sample could give a different successor.
A preferred trajectory is therefore the best one we observed, not a proven optimum, and a failed trajectory may still contain good intermediate choices.
Sampling each repair several times would make the preferences more reliable, but building a tree already costs much more than following one trajectory, since it repairs the program once for every reported type at every state.
The trees also have to be rebuilt when the repair LLM, the agents, or the Verus version change, and applying VeruSteer to another pipeline needs a mapping from its diagnostics to the 35 error types of Table~\ref{tab:error-type-vocabulary}, since any other diagnostic is reported only as \texttt{Other}.
A live pipeline, where every round draws fresh LLM samples, would test these preferences beyond the recorded transitions.

\noindent\textbf{Repair budget.}
We build the trees under a budget of $B=5$ rounds, which keeps the expansion affordable, so the preferences encode which trajectory is better within five rounds.
A choice that does not pay off by round five may still lead to a longer successful proof, and a gain at five rounds may not hold with more rounds.
The window length $L$ cannot make up for this: it sets how many of the latest states the selector reads, not how long a repair it has seen, so a longer window does not teach the selector about longer repairs.
Learning them requires trees collected under a larger budget, whose cost grows with every extra round.

\subsection{Threats to Validity}
\label{sec:threats}

\noindent\textbf{Internal validity.}
We split by task, so no test task appears in training or validation, and we select checkpoints on the validation trees only.
The test trees are not independent samples, since trees of one task share its problem: the 289 test trees with alternative trajectories come from 101 tasks.
Each trained variant is a single checkpoint and the prompt-based selector a single sampled run, so differences between selectors include training and sampling randomness, which we do not measure.
We report the number of trees beside every rate, and read differences of a few trees as indications of the trend rather than precise effects.

\noindent\textbf{Construct validity.}
Verification success means that a program passes the verifier against the supplied specification, not that the specification is adequate.
We therefore audited the verified test leaves against their root programs: all report zero errors, none adds a proof-bypassing construct such as \texttt{assume}, \texttt{admit}, or \texttt{external\_body}, and the few AutoVerus leaves that change a specification element are not among the verifications that VeruSteer gains over the default rule.
The progress score and the number of repair rounds are surrogates as well, since neither measures functional coverage, assurance strength, elapsed time, or human effort.

\noindent\textbf{External validity.}
The evaluation covers 684 of the 762 StarVerus tasks that have repair trees, with two repair pipelines, initial programs from DeepSeek-Chat and Qwen-Coder, and DeepSeek-Chat for all repair rounds, so it does not describe an end-to-end benchmark of generation and repair.
Replay compares every selector on the same recorded transitions, which keeps the comparison between selectors fair, but a fresh LLM call may produce another patch.

\section{Related Work}\label{sec:related}

\subsection{LLM-Assisted Specification and Proof Repair}
\label{sec:related-repair}

AutoVerus~\cite{AutoVerus} generates Verus proof annotations and repairs failures through agents specialized by error type.
VeruSAGE~\cite{VeruSAGE} extends this approach to system-level proofs: its planner uses code, diagnostics, and earlier attempts to choose repair actions, and its candidate selector chooses a patch or reverts to an earlier candidate.
Although these decisions use context and history, the error type to target still follows a preset priority.
VeruSteer learns this target-selection decision, while the underlying pipeline handles repair actions and candidate selection (Section~\ref{sec:background-pipelines}).

StarVerus~\cite{jiang2026starverus} is the closest prior work on repair-target selection.
Its prompted LLM planner analyzes the program, diagnostics, and project context to propose an order over error types.
Adaptive target selection therefore has a precedent.
VeruSteer studies how to learn this decision from repair outcomes: it trains a dedicated selector on preferences between alternative trajectories from the same initial program.
A study of Verus specification failures in Asterinas Vostd~\cite{liu2026beyondbenchmarks} reports obscured root causes, regressions, and stalled repairs, further motivating evaluation across successive repair steps.

Other Verus systems improve repair context or generation.
RAGVerus~\cite{RagVerus} retrieves repository context, and ExVerus~\cite{ExVerus} uses counterexamples to guide repairs of a targeted failure and rank candidate patches.
SAFE~\cite{SAFE} and AlphaVerus~\cite{AlphaVerus} bootstrap generators from verified outputs, while VeriStruct~\cite{VeriStruct} supports generation and repair for data-structure modules.
These techniques improve the context and repair operations available to a pipeline; VeruSteer addresses how to choose among concurrently reported error types.

For other verifiers, Preguss~\cite{PREGUSS,wang2026tale} uses static analysis to identify verification units in C programs and orders their specification synthesis along the call graph.
Its ordering operates over program units before synthesis; VeruSteer's choices respond to changing diagnostics within a repair loop.
Laurel~\cite{Laurel} generates helper assertions for Dafny~\cite{Dafny} using error locations and similar proofs.
Related systems synthesize specifications~\cite{AutoSpec,SpecGen}, loop invariants~\cite{LIG-SE,ACInv,SLD-Spec}, or formalizations of natural-language requirements~\cite{cosler2023nl2spec,NL2ACSL}.
Benchmarks and datasets for verified code and proof generation~\cite{dafnybench,VERINA,VeriCoding,FSTARDATASET,FM-Universe} provide tasks and reference artifacts for training and evaluation.
Our supervision additionally requires alternative error-selection trajectories from a common initial program, which we collect as repair trees on the StarVerus tasks (Section~\ref{sec:setup}).

\subsection{Selection and Search in Verification}
\label{sec:related-selection}

Learned rankers use verification outcomes to prioritize candidates.
iRank~\cite{iRank} learns a contrastive reranker for LLM-generated loop invariants to prioritize valid candidates and reduce verifier calls.
Quarry~\cite{Quarry} learns a pairwise linear difficulty model from recursive proof outcomes to rank Rocq goal decompositions during search.
VeruSteer instead ranks reported error types based on the current state and repair history, using preferences between complete repair trajectories from the same initial program.

Learning also guides multi-step proof construction.
TacTok~\cite{first2020tactok} predicts Coq tactics from proof states and partial scripts, while TacticZero~\cite{wu2021tacticzero} learns tactic and argument selection and backtracking in HOL4.
ReProver~\cite{leandojo} retrieves premises for tactic generation, and ReForm~\cite{ReForm} trains Dafny specification generators with verifier feedback.
VeruSteer uses state, history, and outcome signals to select targets for existing repair agents.
BeeR~\cite{Beer} also studies selection order, combining confidence-based alarm inspection with cluster exploration guided by human feedback.

Online search explores alternative continuations.
AlphaVerus's Treefinement~\cite{AlphaVerus} searches refinement trees for new tasks; counterexample-guided synthesis~\cite{polgreen2020counterexample} and neural search guidance~\cite{kobayashi2025towards} use failed candidates to guide search.
VeruSteer explores alternatives offline for supervision, then follows one repair trajectory without extra repair or verifier calls for selection.

\subsection{Sequential Recommendation and Preference Learning}
\label{sec:related-recommendation}

GRU4Rec~\cite{gru4rec} and SASRec~\cite{kang2018sasrec} predict a user's next item from interaction history using recurrent networks and causal self-attention~\cite{vaswani2017attention}, respectively.
BPR~\cite{rendle2009bpr} learns rankings from pairwise preferences in implicit feedback.
VeruSteer draws on SASRec for sequential encoding and BPR for pairwise ranking, adapting both to error-type selection (Section~\ref{sec:framework}).
Each sequence position combines the program, diagnostics, and preceding selection; available items are the currently reported error types.
Supervision comes from verifier outcomes: a step-wise loss imitates preferred trajectories, and a pairwise loss ranks trajectories from the same initial program.
These established techniques support learning error selection from the outcomes of alternative repair orders.

\section{Conclusion}
\label{sec:conclusions}
We presented VeruSteer, which learns which verifier-error type to repair next from the outcomes of alternative repair trajectories.
Offline repair trees provide preferences among paths from the same initial program, and a sequence model uses the program, verifier feedback, and earlier selections to choose the next target.
The selector operates within the existing repair loop with its specialized agents and verification checks.
Evaluation on stored AutoVerus and VeruSAGE test trees shows that VeruSteer preserves most default verification successes and recovers successful alternatives, with some losses in retention.
It also finds shorter successful paths and improves progress on unsuccessful ones.
The strategy comparisons and ablations show the value of context and trajectory supervision for these choices.
These findings support learning repair order as a separate part of verification automation, allowing existing repair agents to benefit from the outcomes of their alternative executions.


\bibliographystyle{ACM-Reference-Format}
\bibliography{references}

\end{document}